\documentclass[eqsecnum,aps,prd,nofootinbib]{revtex4}

\usepackage[pdftex]{color,graphicx}
\usepackage{enumerate}%
\usepackage{amsmath}%
\usepackage{amsfonts}
\usepackage{verbatim}

\def\be{\begin{equation}}
\def\bea{\begin{eqnarray}}
\def\ee{\end{equation}}
\def\eea{\end{eqnarray}}

\begin{document}

\title{Outgoing gravitational shock-wave at the inner horizon: The late-time limit of black hole interiors}

\author{Donald Marolf\footnote{\tt marolf@physics.ucsb.edu}}

\affiliation{Department of Physics \\ University of California,
Santa Barbara \\ Santa Barbara, CA 93106, USA }

\author{Amos Ori\footnote{\tt amos@physics.technion.ac.il}}

\affiliation{Department of Physics \\ Technion-Israel Institute of Technology \\ Haifa 3200, Israel }

\begin{abstract}
We investigate the interiors of 3+1 dimensional asymptotically flat charged and rotating black holes as described by observers who fall into the black holes at late times, long after any perturbations of the exterior region have decayed.  In the strict limit of late infall times, the initial experiences of such observers are precisely described by the region of the limiting stationary geometry to the past of its inner horizon.
However, we argue that late infall-time observers encounter a null shockwave at the location of the would-be outgoing inner horizon. In particular, for spherically symmetric black hole spacetimes we demonstrate that freely-falling observers experience a metric
discontinuity across this shock, that is, a {\it gravitational shock-wave}.
Furthermore, the magnitude of this shock is at least of order unity.  A similar phenomenon of metric discontinuity appears to take place at the inner horizon of a generically-perturbed spinning black hole.
We compare the properties of this null shockwave singularity with those of the null weak singularity that forms at the Cauchy horizon.
\end{abstract}

\date{\today}

 \maketitle

\tableofcontents

\section{Introduction}
\label{intro}

In Einstein-Hilbert gravity coupled to various matter fields,
the exterior geometry of a 3+1 dimensional asymptotically flat black hole (BH) spacetime typically approaches a stationary solution at late times. Non-stationary perturbations decay both by falling across the horizon and dispersing to infinity, as described by the ringdown of quasi-normal modes followed by power-law tails.  Our purpose here is to explore a corresponding late-time limit of the associated black hole interiors.
We will argue that as far as the observations of
late-infalling physical observers are considered,
the result is well-described by a simple effective geometry which contains the part of the corresponding stationary BH solution to the past of the inner horizon.  However,  the regular inner horizon
is replaced by singular components of two different types:
(i) The {\it ingoing} section of the inner horizon---the Cauchy horizon (CH)---is replaced by a null, weak, curvature singularity, and
(ii) the  {\it outgoing} section of the inner horizon is replaced by an outgoing shock-wave singularity.
The presence of a null, weak, curvature singularity at the CH is a well-known phenomenon since the pioneering works of Hiscock on the Reissner-Nordstrom-Vaidya solution \cite{Hiscock} and of Poisson and Israel on the mass-inflation model \cite{PI} (see also \cite{Ori91}).
It is the second singular component---the outgoing shock-wave singularity---which will be our main concern in this paper.
Our study is motivated in part by the picture of extreme black holes at late times suggested in \cite{extremes} and explored further in \cite{garf}.  This picture agrees with the extreme limit of our results below.

\begin{figure}[ht]
\begin{center}
\includegraphics[scale=.5]{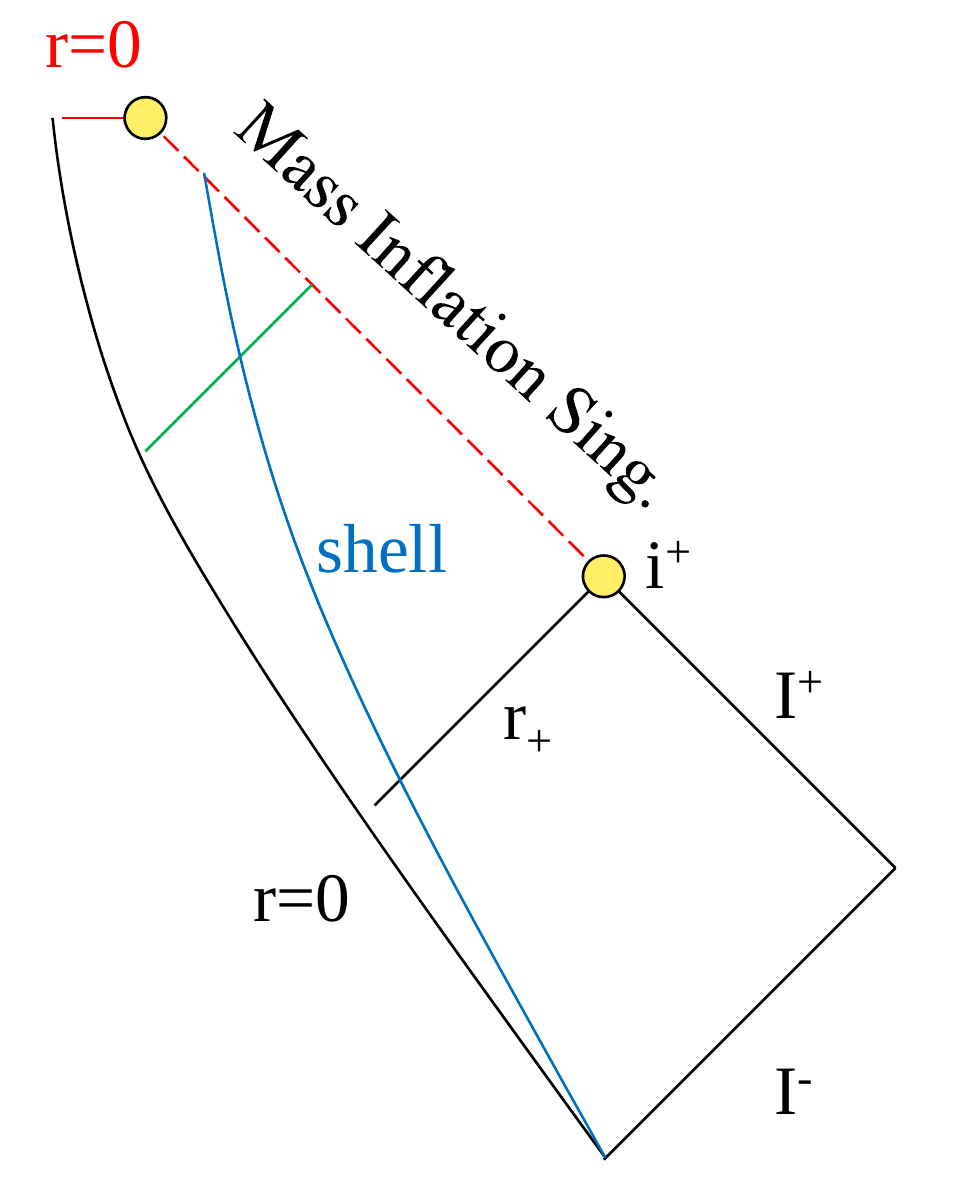}
\includegraphics[scale=.5]{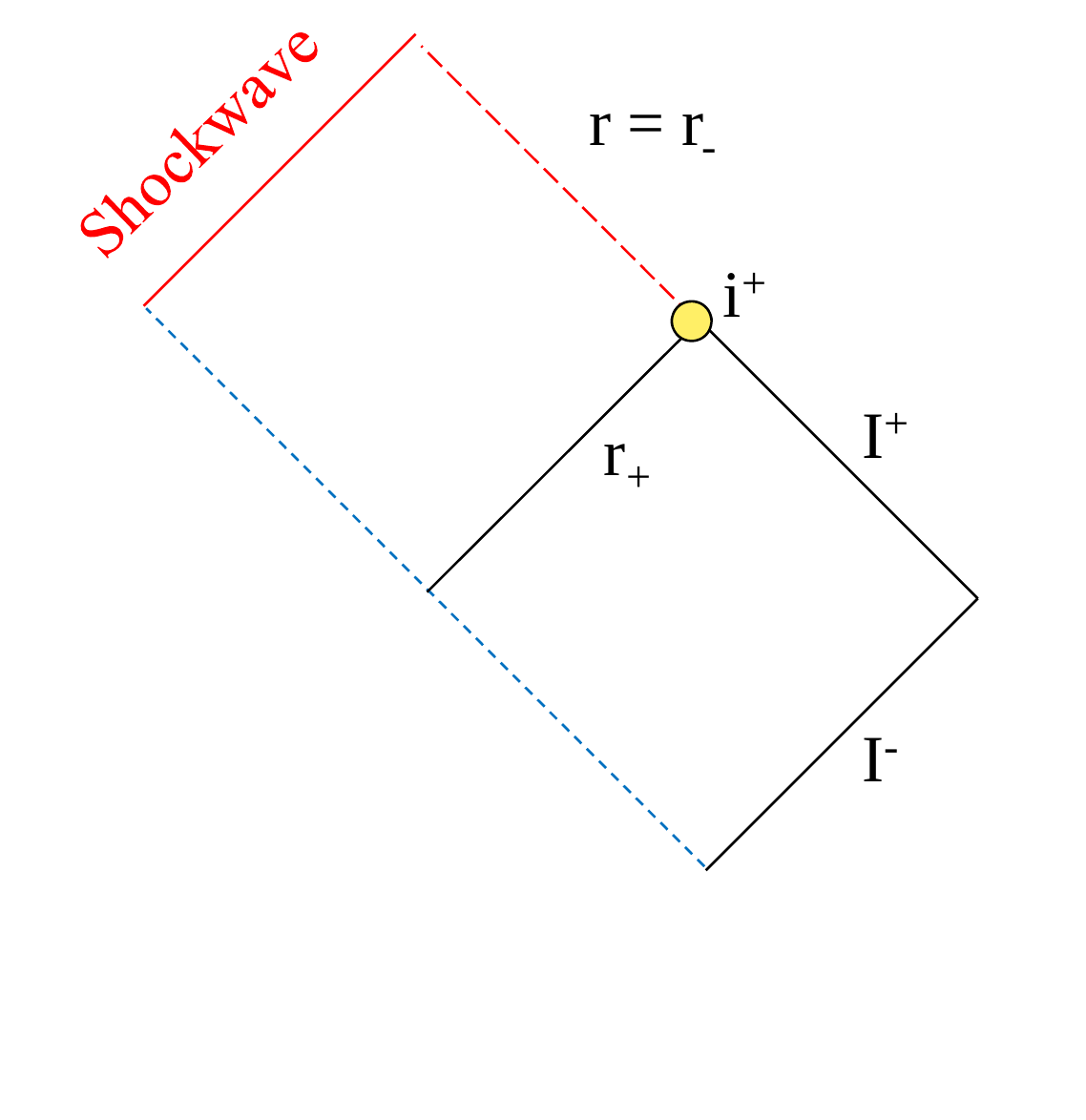}
\caption{ \label{summary}
{\bf Left:} A Reissner-Nordstr\"om black hole formed by collapse of a thin spherical charged shell and subject to spherical perturbations by a massless scalar field.
 Two types of singularities form: A null weak singularity at the CH (the diagonal dashed line) and a spacelike $r=0$ singularity
(the solid horizontal line; though see footnote \ref{insideshell}).
Depending on the strength of the initial perturbation, the spacelike piece of the singularity might also form mcuh earlier.  It may intersect the worldline describing the shell or even entirely remove the would-be outgoing inner horizon.
{\bf Right:} Our proposed effective geometry of the perturbed (but spherically symmetric) collapsing shell spacetime as seen by late-time observers.  It consists precisely of the region of the unperturbed eternal Reissner-Nordstr\"om black hole with $r > r_-$, together with two types of singularities on its future boundary (at $r=r_-$):
The diagonal dashed line at the upper right represents the mass-inflation singularity, a null weak singularity located at the CH.
The diagonal solid line at the upper left represents a null shock-wave singularity, where the metric tensor is effectively discontinuous.
}
\end{center}
\end{figure}

The starting point for our analysis is the large body of literature studying perturbations of the Reissner-Nordstr\"om (RN) and Kerr interiors.
With the assumption of spherical symmetry, these works establish that perturbations transform at least the initial part of the ingoing inner horizon (the CH) of RN into a null curvature singularity often called the mass-inflation singularity. This singularity is weak in the sense that the metric remains continuous at the singularity, though it is not differentiable. In particular, the area-radius $r$ of the spheres is well-defined at this singularity, taking the value $r_- = M - \sqrt{M^2 - q^2}$ near the point marked $i^+$ on the conformal diagram shown in figure \ref{summary} (left) and shrinking toward the future as described by the Raychaudhuri equation, eventually reaching $r=0$.
In the spherical case, at least when the matter content includes a minimally-coupled massless scalar field, it was numerically established \cite{GG,BS, Burko} that when $r$ shrinks to zero the weak null singularity meets a strong spacelike singularity\footnote{
\label{insideshell}
These numerical simulations showed that a transition from a null to a spacelike singularity
occurs in a region which (when mapped to a collapsing-shell spacetime) would correspond to being
{\it outside} the shell (i.e. where the electric field is non-zero).
However, when the initial scalar perturbations are sufficiently weak, the full focussing of the CH to $r=0$ will only occur inside the shell. We strongly expect the formation of a spacelike singularity in this case as well, and have drawn this scenario in figure \ref{summary}.}, along which $r=0$.
This situation is depicted in the left panel of figure \ref{summary}.
Note that curvature scalars diverge at both the above spacelike and null singularities.
 While the establishment of a spacelike singularity is less certain for other forms of matter (e.g. perfect fluids), it is nevertheless expected that all future-directed timelike and null curves inside the black hole will be incomplete due to reaching a curvature singularity of some form.

 Investigations of nonlinearly-perturbed spinning BHs reveal a similar scenario.
Perturbative analyses \cite{Ori92, EMGkerr, OriOsc} again indicate the formation of a null weak scalar-curvature singularity at the CH
 (though this time the singularity is generically oscillatory \cite{OriOsc}, as opposed to the monotonic mass-inflation singularity in the spherical case). The presence of such a null singularity is also supported by an asymptotic
local analysis of the Einstein equations \cite{Brady}, as well as by exact analytical constructions of locally-generic classes of null weak singularity
\cite{Generic}.
 Yet, to the best of our knowledge, no numerical verification of this scenario has yet been carried out in the spinning case.

The results above provide a good (if not yet fully complete) understanding of the
internal structure of generic (charged, rotating, non-stationary) asymptotically
flat, isolated (i.e. non-accreting \cite{accreting}) black holes in classical general relativity   \footnote{See \cite{Thorlacius,OriSemiclassical,Y1,Y2} for attempts to
incorporate semi-classical effects.}.  However, the detailed experiences of any given observer inside such a black hole will in general depend on the process by which the black hole was formed and on the particular perturbations generated. In contrast, we argue below that the spacetime effectively simplifies from the point of view of observers who enter the black hole at late times, which we call late infall-time (or just late-infall) observers.  The simplified spacetime may be described by the simple, stationary BH solution up to the inner horizon---and an effective outgoing shock wave at the outgoing portion of the latter (plus a null weak singularity at the CH).
This structure is depicted in the right panel of figure \ref{summary}.

Due to the key role it plays in our analysis, it is useful to describe the notion of {\it late-infall observers} in more detail.
We start with a simple demonstration (though not necessarily the most precise or most general one) of this concept.
Recall that perturbations outside the BH decay at late time, where of course late means relative to the formation of the black hole and the onset of any significant new perturbations. Therefore, for observers who fall into the BH at sufficiently late time, the exterior geometry will be well approximated by the stationary (and axially-symmetric) BH metric. This allows one to associate specific values of energy $E$ ($=-u_t$) and angular momentum $L$ ($=u_\varphi$) to the late-time geodesics. More importantly, owing to the approximate time-translation of the external geometry, from any ``seed" infalling geodesic $\Gamma_0$, we may construct a one-parameter family $\Gamma$ of similar geodesics, obtained from $\Gamma_0$ by time translation to the future.
(We emphasize that in the present construction the members of $\Gamma$ are {\it exactly} geodesics, all related to $\Gamma_0$ by the approximate time translation.
\footnote{
The geodesics in $\Gamma$ may be constructed as follows. Let $x^\mu=(t,x^i)$ be a set of coordinates for the BH exterior, such that
at late time the associated metric functions are approximately independent of $t$.
We pick a certain point $P_0$ on the ``seed" geodesic $\Gamma_0$ outside the BH,
and time-translate it to the future by a certain amount $\Delta t$. At the new point, which we denote $P'_0$, we set the four-velocity components $u^\mu$ to be numerically the same as those of $\Gamma_0$ at $P_0$. The geodesic $\Gamma'_0$ which emanates from $P'_0$ with those initial four-velocity components $u^\mu$ now becomes a member of the set $\Gamma$.})
Note in particular that all geodesics in $\Gamma$ share (approximately) the same values of $E$ and $L$.
Now, each member of $\Gamma$ is characterized by the parameter $v_{eh}$, namely the value of the advanced time $v$
(Eddington's advanced null coordinate) at which the geodesic crosses the event horizon. For any given seed geodesic $\Gamma_0$, the {\it late-infall observers} are those members of $\Gamma$ characterized by sufficiently large values of $v_{eh}$.

Our main objective in this paper is to characterize the experience of such late--infall observers who move toward the outgoing section of the inner horizon in a generically-perturbed charged (or spinning) BH. We shall see that such observers experience abrupt changes in the amplitude of various perturbing fields -- as well as the metric itself -- while crossing the (would-be) inner horizon. These changes occur within a short proper-time interval whose magnitude decreases exponentially with the infall time $v_{eh}$. For an observer with fixed resolution and  sufficiently large $v_{eh}$ this proper-time interval is so tiny that he experiences the perturbation as an effective shock wave.

The above-mentioned concept of late-infall observers may be generalized, and reformulated in a somewhat more precise manner (though we will not attempt a fully precise definition here).
Consider a continuous one-parameter family of inextendible causal curves labeled by the advanced time $v_{eh}$ at which they cross the event horizon, with $v_{eh}$ taking values in some range $(v_0,+ \infty)$ so that the family includes curves that enter the black hole at arbitrarily late times.  Note that since perturbations outside the black hole decay, the advanced time $v$ can indeed be used as a coordinate along the horizon at sufficiently late times.  We require that, in the limit $v_{eh} \rightarrow \infty$, the part of these curves to the past of the event horizon approaches a {\it stationary} family of such curves in some stationary spacetime; i.e. for which curves with different values of $v_{eh}$ are related by the corresponding time-translation. We further require that, for large $v_{eh}$, the parts of our curves to the future of the event horizon all have the same proper accelerations when expressed using a reference frame parallel-propagated along the world line in terms of the proper time along the worldline after crossing the event horizon.  One may think of this as the assumption that all observers in a given family are equipped with identically pre-programmed rocket ships.  Of course we insist that these reference frames at the event horizon are also related by an approximate time translation at large $v_{eh}$.  For our rather qualitative purposes below it will not be necessary to specify the precise rate at which these limiting behaviors are approached, though some such specification will certainly be needed to derive more precise results.

After a brief review of charged spherical black holes in section \ref{preliminaries}, we study the experiences of late-time observers in stationary spacetimes subject to linear perturbations in section \ref{perturbations}.  Such observers experience no perturbation at all until they would expect to encounter an inner horizon.  However, they effectively encounter a shockwave at the outgoing inner horizon.  While we include a discussion of linearly perturbed Kerr black holes,
 our treatment mainly focuses on the simpler spherically symmetric case.

Non-linear perturbations are addressed in sections \ref{non-linear} and \ref{lateT}.  Here we consider only spherical black holes.  Section \ref{non-linear} addresses the model of a charged BH perturbed by a self-gravitating scalar field.
We show that the experiences of freely-falling late-time observers again agree with those in unperturbed Reissner-Nordstr\"om up to the point where they would expect to encounter the (outgoing section of the) inner horizon.  However, instead of finding a smooth null surface at that point, they effectively encounter a gravitational shockwave at which the metric is discontinuous.  Section \ref{lateT} then gives a heuristic argument that the experiences of more general late-time observers are similar, and in particular that they are described by the effective spacetime shown in the right panel of figure \ref{summary}.
The final discussion in section \ref{disc} describes possible generalizations to rotating black holes and to black holes in any dimensions and the implications for finite-time observers who fall into astrophysical black holes.
We also discuss several aspects of the gravitational shock-wave phenomenon which takes place at the inner horizon.

\section{Preliminaries: Spherical charged black hole}
\label{preliminaries}

\begin{figure}[ht]
\begin{center}
\includegraphics[scale=.5]{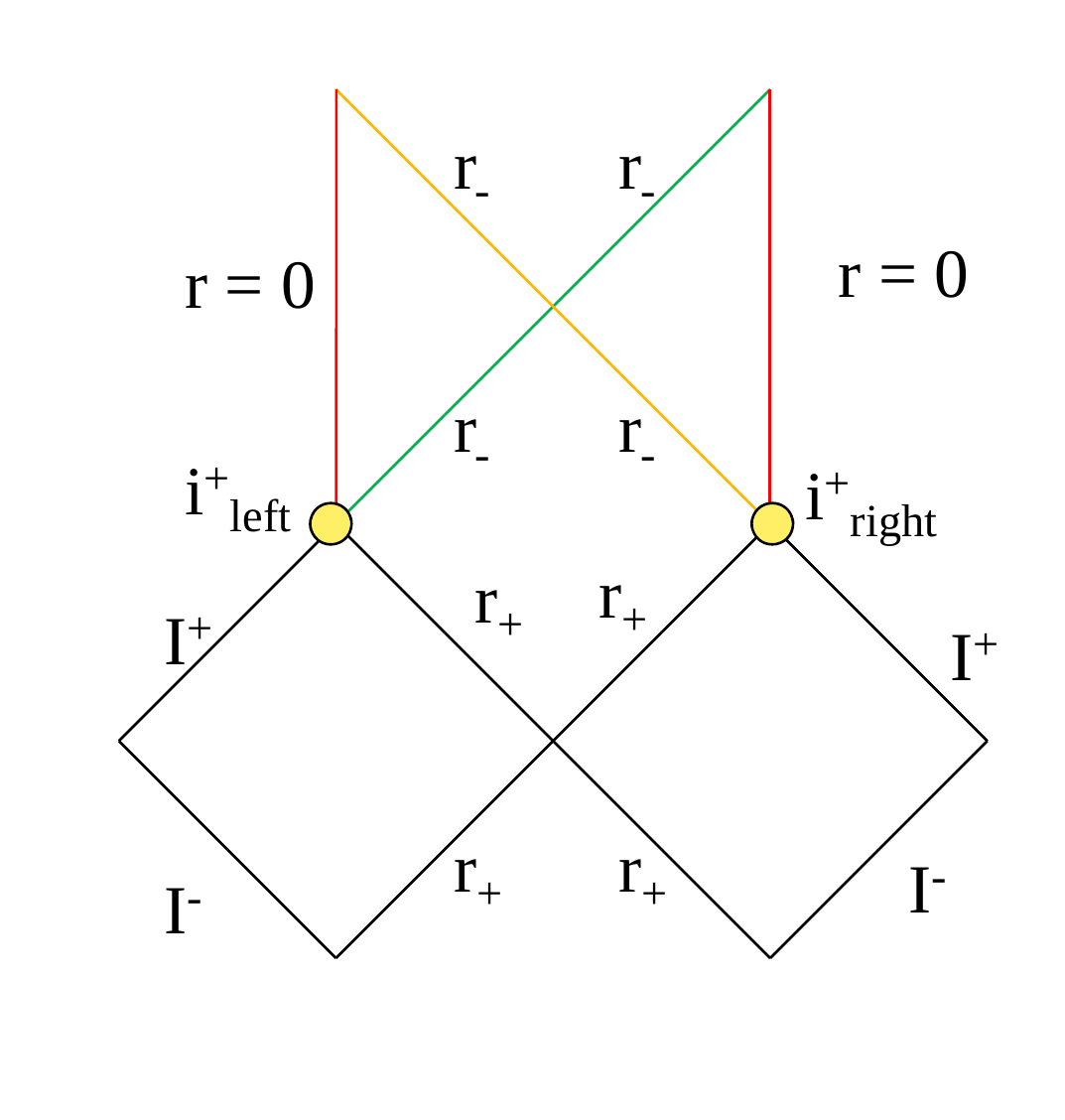}
\includegraphics[scale=.5]{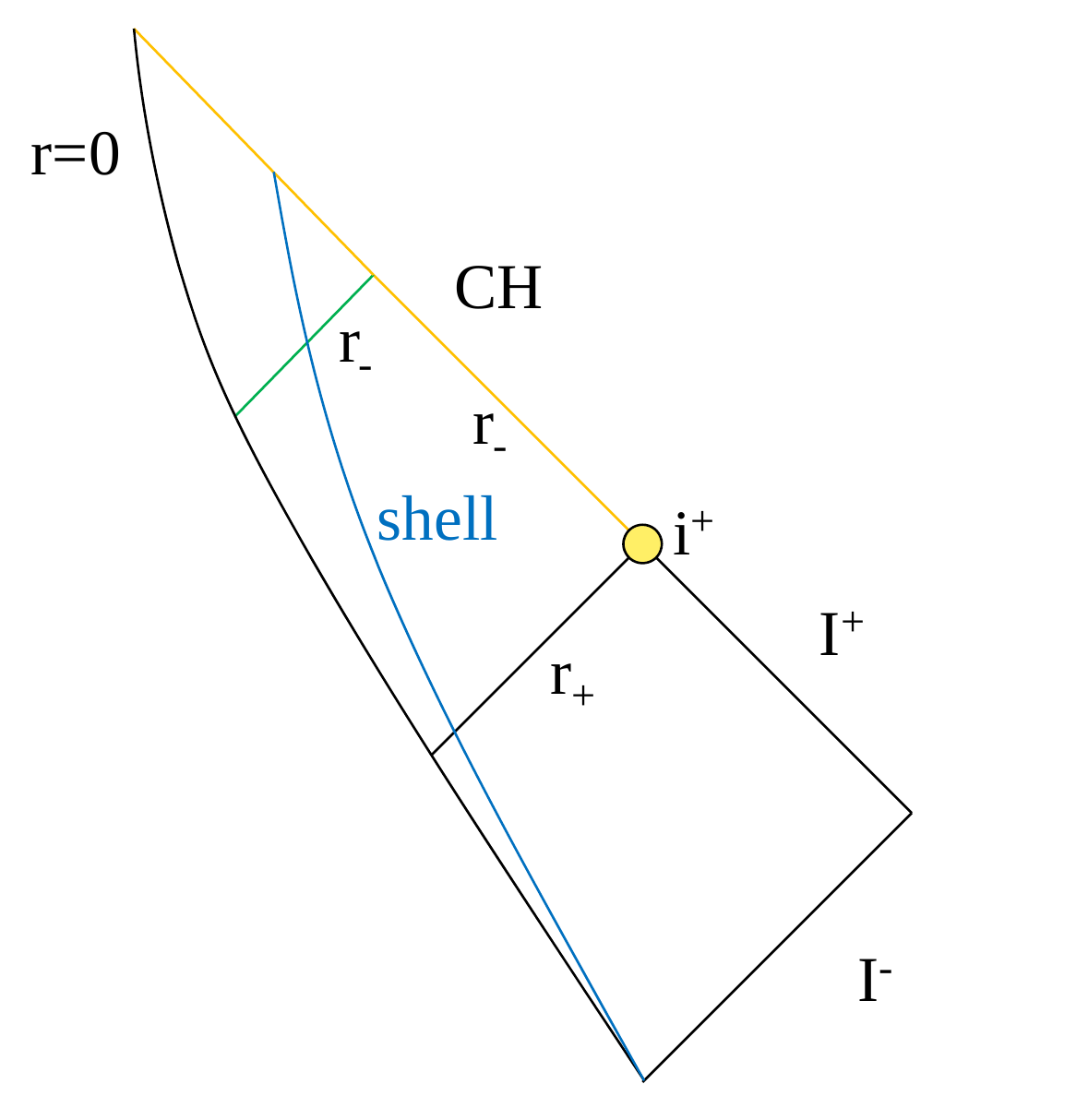}
\caption{\label{RNfig}
{\bf Left:} The eternal Reissner-Nordstr\"om black hole.  The vertical lines are singularities at $r=0$, while the diagonal lines describe copies of asymptotically flat future or past null infinities ($I^+$ and $I^-$), outer horizons ($r_+$), and inner horizons ($r_-$). The full analytically extended solution consists of a periodic vertical array of copies of the regions shown. {\bf Right:} A conformal diagram for the spacetime of a collapsing (but otherwise unperturbed) spherically symmetric massive charged thin shell.  The $r=0$ curve is a regular origin in the flat region inside the shell.  The Cauchy horizon is also indicated.}
\end{center}
\end{figure}

The RN solution is the unique spherically-symmetric electrovac geometry.
In Schwarzschild coordinates $(t,r,\theta,\varphi)$ it takes the form
\begin{equation}
ds^2 = -F dt^2+F^{-1} dr^2 + r^2 d\Omega_2^2 \, ,
\label{RN}
\end{equation}
where $F\equiv 1-2M/r+q^2/r^2$, and $d\Omega_2^2\equiv d\theta^2+sin^2(\theta) d\varphi^2$ is the unit two-sphere.
Throughout this paper we shall consider the non-extreme black-hole (BH) case, namely $0<|q|<M$. In this case $F(r)$ vanishes at two $r$ values, $r_\pm\equiv M\pm\sqrt{M^2-q^2}$. The larger root $r_+$ corresponds to the event horizon, and the smaller one $r_-$ to the inner horizon.
Figure \ref{RNfig} (left)
depicts a part of the Penrose diagram of the eternal, analytically-extended, RN geometry.
Note that the inner horizon has two separate portions -- the two intersecting null lines denoted ``$r_{-}$".

Later we shall also consider the spherically-symmetric spacetime of a charged collapsing thin shell. The geometry outside the shell is described by the Reissner-Nordstrom (RN) metric (\ref{RN}). Inside the shell the geometry is flat, i.e. Minkowski.
Figure \ref{RNfig} (right)
displays this hybrid spacetime. In this non-eternal BH spacetime the ingoing portion of the inner horizon is a Cauchy horizon (CH).
The diagram only  displays the globally-hyperbolic piece of the spacetime---namely, the region up to the Cauchy horizon (the extension beyond the CH will not concern us in this paper).

Let us examine the free-fall orbits of observers who jump into the BH.
Throughout this paper we assume, as usual (and without loss of generality) that the motion is confined to the equatorial plane.
These geodesic orbits are characterized by two constants of motion:
the ``Energy" $E\equiv -u_t>0$, and the angular momentum $L\equiv u_\varphi$.
They satisfy the radial equation
\footnote{At a certain value of $r$, $r=r_{bounce}<r_-$, $dr/d\tau$ flips its sign (a phenomenon known as ``gravitational bounce"). In this paper, however, we focus on the behavior of orbits up to their first approach to $r=r_-$, hence $dr/d\tau<0$.}
\begin{equation}
\frac{dr}{d\tau}=-\sqrt{E^2-(1+L^2/r^2)F(r)} .
\end{equation}
Of particular importance will be the behavior of these worldlines in the neighborhood of $r=r_-$, where the last result becomes
\begin{equation}
\frac{dr}{d\tau}\cong -E.
\label{Delta_tau}
\end{equation}

In the analysis below it will be useful to express the RN metric in double-null coordinates.
A particularly useful form is given by the Eddington coordinates
\begin{equation}
u\equiv t-r_* \, , \,\, \,v\equiv t+r_*,
\label{uvDEFINE}
\end{equation}
where $r_*$ is the tortoise coordinate defined by
\begin{equation}
dr/dr_*=F(r).
\label{tortoise}
\end{equation}
The line element then becomes
\begin{equation}
ds^2 = -F(r)dudv+ r(u,v)^2 d\Omega_2^2 \, ,
\label{Eddington}
\end{equation}
where $r$ is to be regarded as a function of $u$ and $v$,
determined (implicitly) by setting $r_*=(v-u)/2$ in Eq. (\ref{tortoise}).
Notice that in the region which will mostly concern us here---the domain $r_-<r<r_+$ --- $u$
is {\it past-directed} (this choice of sign simplifies many of the expressions below).

Note that $r_*\to +\infty$ at $r=r_-$, implying that either $u$ or $v$ must diverge there.
It follows that at the ingoing section of the inner horizon $v\to \infty$ (as $u$ is regular), whereas at the outgoing section $u\to -\infty$ (and $v$ is regular).

Since $F$ vanishes at the inner horizon, in its neighborhood we may approximate $F(r)\cong -2\kappa (r-r_-)$, where $\kappa\equiv -(1/2)(dF/dr)_{r=r_-}$ (note that $\kappa>0$). It then follows\footnote{Eq. (\ref{tortoise}) defines $r_*$ up to an integration constant. We use this freedom and choose the convenient pre-factor $M$ in the right-hand side of Eq. (\ref{Delta_r}),  which fixes this arbitrary constant.}
from Eq. (\ref{tortoise}) that near the inner horizon
\begin{equation}
r-r_- \cong M e^{-2\kappa r_*}.
\label{Delta_r}
\end{equation}
That is, $r_*$ diverges logarithmically (in $r-r_-$) at the inner horizon.

The metric (\ref{Eddington}) is singular at the inner horizon (where $\det (g)$ vanishes). To remove this coordinate singularity we define the inner-horizon's Kruskal-like coordinates
$U\equiv -e^{\kappa u}$, $V\equiv -e^{-\kappa v}$.
With this choice of signs, both $U$ and $V$ are future-directed.
Note that $V$ ($U$) vanishes at the ingoing (outgoing) section of the inner horizon.
The line element now reads
 $2g_{UV}dUdV+ r^2 d\Omega_2^2$.
We shall not need here the specific form of the metric functions $r(U,V)$ and
$g_{UV}(U,V)=g_{UV}(r)$.
We shall just note that both functions are regular. Furthermore, $g_{UV}$ turns out to be a smooth function of $r$ which (unlike $F$)
 is nonvanishing at $r_-$. As a consequence, the Kruskal metric is
 regular at the inner horizon.

\section{Simple examples of late-time perturbations:
Linearized fields}
\label{perturbations}

In this section we consider several types of
linear perturbations, and explore how these perturbations are experienced by late-time infalling observers.

We shall start our analysis by addressing the simplest type of perturbation, namely a purely-outgoing, spherically-symmetric, test scalar field.
Then we shall proceed to consider more realistic types of
linear perturbations, deferring non-linear perturbations to the following sections.

\subsection{
Simplest example: monotonic, outgoing, test scalar perturbations}
\label{monotonic}

Consider a free, massless, minimally-coupled, test scalar field $\phi$ on the RN background.
The general behavior of a scalar field of this type inside the BH will be discussed in the next subsection.
Here we focus on a  spherically-symmetric scalar field,
restricting our attention to the very neighborhood of the Cauchy horizon.
As it turns out, in this region
the field becomes purely outgoing, namely
\footnote{This simple form
follows from the behavior of
perturbation fields inside BHs, which we discuss in the next subsection: In the very neighborhood of the inner horizon, the outgoing and ingoing modes effectively decouple [see Eq.(\ref{nearIH})]. Furthermore, the $v$-dependent component of $\phi$ decays as $v^{-n}$ at large $v$, leaving us with only the $u$-dependent component.}
 $\phi\cong P(u)$.
In our first example we shall further assume, for simplicity, that $P(u)$ vanishes until a certain value $u_1$; then it grows monotonically up to $u=u_2$, where it reaches its final value $P_f$ (and remains constant afterward).
While this type of function $P(u)$ is certainly oversimplified,  it  transparently demonstrates the mechanism responsible for the shock-wave formation. \footnote{Essentially the same analysis will apply in the more general
(and more realistic) case, in which the outgoing field starts
right after the event horizon,
and may also continue its growth beyond the (outgoing portion of the) inner horizon.
Also the function $P(u)$ needs not be monotonic
(see discussion below).
Here we picked a monotonic function $P(u)$
only for the sake of conceptual simplicity.}
Both null lines $u=u_{1,2}$ are assumed to reside in the internal range $r_-<r<r_+$.
We shall now explore the behavior of $\phi$ along the worldline of the infalling observer, as a function of proper time $\tau$---focussing our attention on late-time observers.

A free-falling observer in a RN spacetime crosses the inner horizon $r=r_-$ through its {\it outgoing} section (where $v$ is finite). Let us denote the values of $v$ as the observer crosses the event and inner horizons by $v_{eh}$ and $v_{ih}$, respectively. Since we are interested in late-time observers we shall assume $v_{eh}\gg M$.
\footnote{
Within the shell-collapse model we further assume that $v_{eh}$ is sufficiently large that the orbit is confined to the shell's exterior throughout the range $r>r_-$.
Also, in the case $u_1>0$ we further demand
$v_{eh}\gg M+u_1$, such that $r_*\gg M$ as $u=u_1$ is approached.}
Along a timelike worldline $v$ increases monotonically, hence in the range between the two horizons $v_{eh}<v<v_{ih}$.

Obviously, $v_{eh}$ and $v_{ih}$ depend on the infalling geodesic. However, for fixed parameters $E$ and $L$, the difference
$\Delta v\equiv v_{ih}-v_{eh}=\Delta v(E,L)$
will be independent of the infall time,
owing to the time-translation symmetry of the RN metric.  $\Delta v$ is typically of order $M$
(we assume here that $E$ is of order unity,
i.e. not too large or too close to zero, and $|L|$ is $\sim M$ or smaller; and similarly the BH is not too close to Schwarzschild or to extremality.)

Next we evaluate
the proper times $\tau_{1,2}$ at the two events where the worldline intersects the null lines $u=u_{1,2}$.
For convenience we set $\tau=0$ at the worldline's intersection with the inner horizon. From the above assumptions (in particular $v_{eh}\gg M$) it immediately follows that $r_*\gg M$ --- and hence $r\cong r_-$ --- throughout the portion $u_1>u>u_2$ of the worldline. We can therefore use Eqs. (\ref{Delta_tau}) and (\ref{Delta_r}) in this domain to obtain
\begin{equation}
\tau \cong -E^{-1}(r-r_-)\cong -M E^{-1} e^{-2\kappa r_*}.
\label{}
\end{equation}
Substituting $r_*=(v-u)/2$ we find for $\tau_{1}$ and $\tau_{2}$
\begin{equation}
\tau_{1,2} \cong -M E^{-1} e^{\kappa(u_{1,2}-v_{1,2})},
\label{tau12}
\end{equation}
where $v_{1,2}$ respectively denote the value of $v$ at the worldline's intersections with $u=u_{1,2}$.

Since $v_{1,2}>v_{eh}$, we readily find
\begin{equation}
|\tau_{1,2}|< (M E^{-1} e^{\kappa u_{1,2}})  e^{-\kappa v_{eh}},
\label{bound}
\end{equation}
hence the difference $\Delta \tau_{12}\equiv \tau_2-\tau_1>0$ is bounded by
\begin{equation}
\Delta \tau_{12}< (M E^{-1} e^{\kappa u_1})  e^{-\kappa v_{eh}}.
\label{DifBound}
\end{equation}

The last inequality tells us at once that the late-time infalling observers will see a scalar-field profile $\phi(\tau)$ which rises from zero (at $u_1$) to its maximal value $P_f > 0$ (at $u_2$) within an arbitrarily short proper-time interval, $\propto \exp(-\kappa v_{eh})$.
 For a sufficiently large $v_{eh}$, this proper-time interval will presumably be unresolved by an observer with fixed sophistication.
 The large-$v_{eh}$ observers will thus experience the scalar perturbation as a sharp shock-wave of finite amplitude.   Note that in terms of the proper time of these observers, the shock wave is detected effectively at $\tau=0$ [the limit $v_{eh}\to\infty$ of
Eq. (\ref{bound})]---namely, just at the crossing time of the (outgoing portion of the) inner horizon.

It should be emphasized that the two outgoing null lines $u= u_1$ and $u=u_2$ need {\it
not} be close to the (outgoing section of the) inner horizon in any sense. To
further clarify this point, consider the intersection of these two outgoing rays
with an ingoing null geodesic $v=const\equiv v^{col}$ located just after the
collapse. (To be more specific, within the shell-collapse scenario we may choose
$v^{col}$ such that it passes through the intersection point of the collapsing shell
with $u=u_2$.)  Let us denote the $r$ values at the two intersection points of
$v=v^{col}$ with $u_1$ and $u_2$ by $r^{col}_1$ and  $r^{col}_2$ respectively. Our
point here is that $r^{col}_1$ and  $r^{col}_2$ need {\it not} be close to $r_-$;
rather, they can take values anywhere in the range $r_ - < r^{col}_2 < r^{col}_1
<r_+$. Still, since $r_*=(v-u)/2$, at a sufficiently large $v$ both these $u=const$
lines will necessarily attain $r_*\gg M$ values, and hence $r$ values arbitrarily
close to $r_-$.

It may be instructive to complement Eq. (\ref{DifBound}) by an actual estimate of  (rather than just a bound on) $\Delta \tau_{12}$, for late infalling observers.
Based on Eq. (\ref{tau12}) we express $\Delta \tau_{12}$ as
 \begin{equation}
\Delta \tau_{12} \cong \frac{M}{E}
\left(e^{\kappa[u_{1}+(v_{ih}-v_{1})]} - e^{\kappa[u_{2}+(v_{ih}-v_{2})]}  \right)
 e^{-\kappa v_{ih}} .
\label{DELTAtau12}
\end{equation}
Consider now the quantity $v_{ih}-v_{1}$. For fixed parameters $E$ and $L$, $dv/d\tau$ is a well-defined function of $r$ (independent of $v_{eh}$). In the late-infall limit, $r\to r_-$ when the orbit reaches $u_1$.
The quantity $dv/d\tau$ at $u_1$ similarly
approaches
its inner-horizon value, which (for $E>0$) turns out to be a definite finite number, $(1+L^{2}/r_-^{2})/(2E)$.
In addition, from Eq. (\ref{bound}) it follows that  $\tau_1\to 0$
in the late-infall limit, and thus that $v_{ih} - v_1 \rightarrow 0$.
Obviously exactly the same argument applies to $v_{ih}-v_{2}$ as well.
Using $v_{ih}-v_{1,2}\to 0$ and $v_{ih}=v_{eh}+\Delta v$ in Eq. (\ref{DELTAtau12}), we thus obtain
\begin{equation}
\Delta \tau_{12} \cong \left[\frac{M}{E} (e^{\kappa u_1}-e^{\kappa u_2})e^{-\kappa \Delta v}\right]  e^{-\kappa v_{eh}}.
\label{DifEstimate}
\end{equation}
(Note that the pre-factor in squared brackets is independent of the infall time $v_{eh}$.)

\begin{figure}[ht]
\begin{center}
\includegraphics[scale=1]{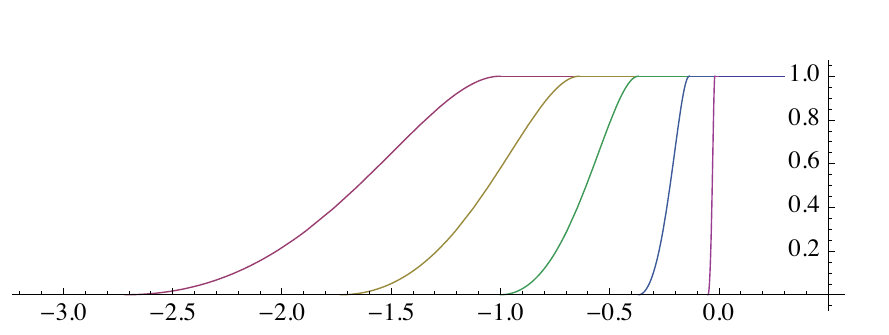}
\caption{\label{simple}
The typical behavior of $\phi$ as a function of the observer's proper time is depicted for several values of $v_{eh}$.
}
\end{center}
\end{figure}

The typical behavior of $\phi$ as a function of the observer's proper-time is depicted in Fig. \ref{simple} for several values of infall time $v_{eh}$.

 Summarizing, the late-time infalling observers will experience the scalar perturbation as a sharp shock wave of finite amplitude ($\Delta \phi=P_f$), with effectively vanishing rising time ($\propto e^{-\kappa v_{eh}}$)---located just at the outgoing section of the inner horizon.

 So far we have considered the case of a (near-CH) outgoing field $\phi\cong P(u)$ which varies monotonically from $P=0$ at $u=u_1$ until $u=u_2$ where  it saturates at $P=P_f$.
 The extension of the argument to a more general function $P(u)$ (not necessarily monotonic; and not necessarily one with well-defined final saturation value) is straightforward: It is sufficient to assume that in the near-CH region,
 $P(u)$  varies over a certain range $u_1>u>u_2$, from $P_1$ at $u_1$ to some $P_2 \neq P_1$ at $u_2$.
 Then, the above analysis yields that for a late-infall observer the finite jump
$\Delta \phi = P_2-P_1\neq 0$
 in the value of $\phi$ will take place within an extremely small proper-time interval, which decreases as $e^{-\kappa v_{eh}}$.
 For an observer of given sophistication, this exponentially small proper time will become unresolvable at sufficiently large $v_{eh}$. Hence, a late-infall observer will again experience a kind of effective shock-wave phenomenon: a finite change in the field, within an effectively-vanishing proper-time interval.
 \footnote{A ``classic" shock wave typically contains the following three components: (i) a steady pre-shock value, (ii) a (different) steady post-shock value, and (iii) a sharp transition between the two, through a transition region of
  arbitrarily-small width.
 Here we use a somewhat generalized notion of shock wave, because we do have the component (iii), and effectively also (i), but we don't necessarily have (ii).
 }

\subsection{More realistic linear perturbations}
\label{linear}

We turn now to consider more realistic perturbations of a spherical charged BH. We still assume that in the region of interest the perturbations are small and can therefore be treated linearly. Again, we would like to explore how these perturbations will be experienced by late-time infalling observers.

 \subsubsection{Non-spherical test scalar field}
 \label{nonspherical}

The linearized perturbations---both outside and inside the BH---may conveniently be toy-modeled by a minimally-coupled massless scalar field $\phi$, satisfying $g^{\alpha\beta} \phi_{;\alpha\beta}=0$ \cite{Price}.
The perturbation field may be decomposed in spherical harmonics $Y_{lm}$ in the usual way,
 \begin{equation}
\phi(x^\mu)=\sum_{lm}Y_{lm}(\theta,\varphi) \phi_{lm}(r,t).
\label{Ylm}
\end{equation}
The individual perturbation modes $ \phi_{lm}$ all satisfy a hyperbolic partial differential equation of the form
 \begin{equation}
 \frac {d^2 \psi_{lm}}{dr_*^2}- \frac {d^2 \psi_{lm}}{dt^2} =V_l(r) \psi_{lm} ,
\label{ScalarEquation}
 \end{equation}
 with $\psi_{lm}\equiv r \phi_{lm}$, and with an $l$-dependent effective potential given by
  \begin{equation}
  V_l(r) =F \left[ \frac{l(l+1)}{r^2} +\frac{2(M-q^2/r)}{r^3} \right]  .
\label{ScalarPotential}
 \end{equation}

 One then finds that outside the BH all modes $\phi_{lm}$ decay to zero at late time \cite{Price}.  This decay typically proceeds in two stages: First is the stage of ``quasi-normal ringing" (i.e. exponentially-damped oscillations). Subsequently, after the ringing has damped, the late-time perturbations are dominated by inverse-power tails. All modes $\phi_{lm}$
decay (along worldlines of constant $r$) as $~t^{-n}$, with $n=2l+3$ (or $n=2l+2$ for initially-static multipoles)
\cite{Bicak, Bicak2}---the same inverse-power form as in the Schawarzschild case \cite{Price}.
Throughout the rest of the paper we shall
focus our attention on the inverse-power tails
(rather than the quasi-normal ringing),
because it is this component which eventually dominates at late time.

Investigations of the dynamics of linearized perturbations inside the BH reveal a behavior which parallels the external dynamics in many respects, though there also are some important differences. The infalling
power-law tails lead to a similar inverse-power decay inside
the BH. Thus, along lines of constant $r$ (between the event and inner horizons), the perturbations still decay as $~t^{-n}$ \cite{Gursel}. Note, however, that this time $t$ is a spacelike coordinate, so what we face here is a {\it spatial}
rather than temporal decay. (For a discussion of this issue see Ref. \cite{OriGRG}.)

Being proportional to $F$, the effective potential $V_l(r)$ vanishes at $r_-$ (exponentially in $r_*$, as it also does at $r_+$). As a consequence,
in the neighborhood of the inner horizon the perturbations take the simple form of a superposition of outgoing and ingoing modes, that is,
 \begin{equation}
\phi_{lm}\cong P_u(u)+P_v(v).
\label{nearIH}
\end{equation}
Hereafter we omit the sub-indices $l,m$ from the $P$-coefficients for brevity.

In the asymptotic region $u,v\gg M$, both functions $P_u$ and $P_v$ admit simple inverse-power forms:
\begin{equation}
 P_u(u)\cong a u^{-n}\, , \, \,  P_v(v)\cong b v^{-n},
 \label{tails}
\end{equation}
again with $n=2l+2$ or $n=2l+3$ (depending on the presence or absence of initial static multipole).
The coefficients $a,b$ are determined by the scattering problem, and are given in Ref. \cite{Gursel} (see also \cite{OriRNco}).

We consider now a late-infall observer, and examine how this observer will record the scalar perturbations, focusing on a particular multipole $\phi_{lm}$.
We restrict our attention to the orbit's section between two hypersurfaces $u=u_1$ and $u=u_2$ (both reside in the domain between the event and inner horizons,
and satisfy
$u_2<u_1$).
We focus on the quantity
$\Delta \phi_{lm}\equiv \phi_{lm}(\tau_2)-\phi_{lm}(\tau_1)$,
where as before, $\tau_{1,2}$ respectively denote the proper times when the observer crosses the two hypersurfaces $u=u_{1,2}$.

For a sufficiently late infall time, the observer will reach $u=u_1$
only when the orbit is already in the near-CH region, where Eq. (\ref{nearIH}) applies. Therefore,
\[ \Delta \phi_{lm} \cong [P_u(u_2)-P_u(u_1)]+[P_v(v_2)-P_v(v_1)] , \]
where $v_{1,2}\equiv v(\tau_{1,2})$.
Furthermore, since $P_v\propto v^{-n}$ at large $v$ (and  $v_{2}>v_{1}>v_{eh}$), it follows that the second term in squared brackets vanishes at the late-infall limit. Thus, for a sufficiently large $v_{eh}$ the expression for $\Delta \phi_{lm} $ simplifies to
\begin{equation}
\Delta \phi_{lm} \cong P_u(u_2)-P_u(u_1)  .
\label{DeltaPu}
\end{equation}
For generic choice of $u_1,u_2$ the RHS is non-vanishing. (This is most explicitly verified in the case where $u_{1,2}$ are both in the early domain, $u\gg M$, where $P_u\propto u^{-n}$.
\footnote{Yet, a more appreciable $\Delta \phi_{lm}$ should be achieved when $u_2$ is not
$\gg M$ (and $u_1$ is not too close to $u_2$). }
) Yet, the proper-time difference $\Delta \tau_{12}= \tau_2-\tau_1$ is still given by Eq. (\ref{DifEstimate}), namely, $\Delta \tau_{12}\propto  e^{-\kappa v_{eh}}$. Thus, for sufficiently late infall, the observer will watch a finite jump $\Delta \phi_{lm}$ in the field component $\phi_{lm}$, during an effectively vanishing (i.e. physically non-resolvable) proper-time interval.

    \subsubsection{Linear electromagnetic and gravitational perturbations}

    In the RN spacetime, owing to the presence of electric field, the (electrovac) gravitational and electromagnetic perturbations are mutually coupled already at the linear level.
 Still, one can write decoupled field equations for a pair of combined electromagnetic/gravitational variables (i.e. two specific linear combinations of the electromagnetic and gravitational variables) \cite{chandra}.
 In particular, based on a formalism developed by Moncrief  \cite{ Moncrief}, Gursel et al. \cite{Gursel2} constructed such a pair of electromagnetic/gravitational field variables $R_\pm^{lm}$ (for the various angular modes $l,m$).
    These variables satisfy the decoupled equations
   \begin{equation}
 \frac {d^2 R_\pm^{lm}}{dr_*^2}- \frac {d^2 R_\pm^{lm}}{dt^2} =V_\pm^l(r) R_\pm^{lm} ,
\label{EMGequation}
 \end{equation}
with the effective potential
  \begin{equation}
 V_\pm^l(r) =F \left[ \frac{l(l+1)}{r^2} +\frac{-3M\pm C+4q^4/r}{r^3} \right]  ,
\label{EMGPotential}
 \end{equation}
 where $C=[9M^2+4q^2(l-1)(l+2)]^{1/2} $. In turn, the gravitational and electromagnetic perturbations may be recovered from the fields $R_\pm^{lm}$ (by certain linear combinations of the latter fields and their derivatives) \cite{chandra}.

The potentials $V_\pm^l(r)$ are again $\propto F$ and therefore vanish at the two horizons, hence near the CH each of the fields $R_\pm^{lm}$ is decomposed into free  ingoing and outgoing components, i.e. $P_u(u)+P_v(v)$, as in
Eq. (\ref{nearIH}). Furthermore, the leading-order behavior of these two components in the early-CH domain $u,v\gg M$ was found \cite{Gursel2} to be of the same
form as in the scalar case, i.e. Eq. (\ref{tails}).
As before, our objective is the proper-time variation of the perturbation fields, as recorded by a late-infall observer. The analysis of the previous subsections apply here with no modifications, implying that the variables $R_\pm^{lm}$ undergo finite variations within proper-time intervals $\propto e^{-\kappa v_{eh}}$.
The gravitational and electromagnetic perturbations (constructed from the variables $R_\pm^{lm}$ and their derivatives) are likely to yield a similar structure of an effective shock wave.

\subsubsection{Linear perturbations  in Kerr spacetime }

In this subsection we very briefly
address the case of a linearly-perturbed Kerr BH.
The latter's internal structure is known to be similar in many respects to that of a RN BH.
In particular there are two horizons, an event horizon located at $r=r_+$ and an inner horizon at $r=r_-$, where $r_\pm\equiv M\pm\sqrt{M^2-a^2}$.
Not surprisingly, we find that the effective shock-wave phenomenon takes place in the Kerr case as well.

We focus here on the {\it gravitational} perturbation, which is apparently the perturbation field of greatest physical relevance here.
(Note that the Kerr background---unlike the electrovac RN background---admits pure vacuum gravitational perturbations.)

The behavior of late-time gravitational perturbations inside a Kerr BH has been analyzed using two different formulations:
(i) Analysis of metric perturbations (MP)
$h_{\alpha \beta}\equiv g_{\alpha \beta}-g_{\alpha \beta}^{(kerr)}$ \cite{Ori92},
(ii) Analysis \cite{EMGkerr,OriOsc} of the evolution of the Teukolsky variables \cite{Teukolsky}
$\psi_0$ and $\psi_4$. Both analyses examined the late-time gravitational perturbations, employing the so-called {\it late-time expansion} \cite{OriGRG,OriKERRco,EMGkerr}.
\footnote{Ref. \cite{Ori92} also considered nonlinear metric perturbations, that is,
higher-order terms in the nonlinear perturbation expansion (which turned out, however, to be negligible compared to the linear metric perturbations).
Ref. \cite{EMGkerr} also considered linear electromagnetic perturbations.
In this section, however, we only consider linear gravitational perturbations.}
They both focussed on the near-CH behavior, and led to similar (and mutually-consistent) results.

For the sake of the present analysis the key result
may be summarized as follows:
Near the CH (that is, $v-u\gg M$), the linear MP decouple into a superposition of outgoing and ingoing components, namely
\footnote{In the Kerr case we still define $v\equiv t+r_*$ and $u\equiv t-r_*$, with $r_*$ now defined through $dr/dr_* =(r^2-2M r+a^2)/(r^2+a^2)$ (where $t$ and $r$ are the Boyer-Lindquist time and radial coordinates).}
\begin{equation}
h_{\alpha \beta} \cong h^{u}_{\alpha \beta}(u)+h^{v}_{\alpha \beta}(v) .
\label{htails}
\end{equation}
Furthermore, $h^{v}_{\alpha \beta}$ decays at $v\gg M$ as an inverse power of $v$,
hence for a late-infall observer $h_{\alpha \beta} \cong h^{u}_{\alpha \beta}(u)$ near the CH.

In the Kerr background (unlike the RN case), infalling timelike geodesics may intersect the inner horizon $r=r_-$ either at its ingoing or outgoing section. Here we shall focus on those orbits intersecting the {\it outgoing} section (these include, for example, all infalling geodesics with
positive $E$ and $a L\leq 0$).
The effective shock-wave phenomenon will only occur for this class of orbits.

Consider now an observer which falls into the Kerr BH (and heads towards the outgoing section of the inner horizon), at the late-infall limit. As before, we focus on the observer's history while moving between $u=u_1$ and $u=u_2$.
Just like in the RN case, both proper-times $\tau_{1,2}$ scale at the late-infall limit as
$\propto e^{-\kappa v_{eh}}$ --- and so does their difference  $\Delta \tau_{12}$.
Here $\kappa = \kappa (M,a)$ is a certain positive constant, the inner-horizon's surface gravity.
Again, we find a finite jump
$\Delta h_{\alpha \beta} = h^{u}_{\alpha \beta}(u_2)-h^{u}_{\alpha \beta}(u_1)$ in the metric,
within an effectively-vanishing proper time $\propto e^{-\kappa v_{eh}}$ ---
namely, an effective gravitational shock wave.

On the other hand, geodesics with $aL>2EMr_- $
will intersect the {\it ingoing} section of $r=r_-$.
In the late-infall limit, these observers will hit the CH at its past boundary.

    \subsection{Interpretation in terms of late-time Eddington frames}

We shall provide here a simple interpretation of the effective shock-wave phenomenon derived above.
For concreteness and simplicitly we present the explicit argument only for the RN case, but it applies to the Kerr case as well.

The line element  (\ref{Eddington}) preserves its form under a coordinate transformation of the form
\begin{equation}
u\to\tilde u=u-\delta \, ,\, \, v\to\tilde v=v- \delta \, ,
\label{EDshift}
\end{equation}
where $\delta$ is any constant (this invariance reflects the time-translation symmetry of RN). We shall refer to different sets of Eddington coordinates ---corresponding to different choices of $\delta$---as different
{\it Eddington frames}
(this terminology is borrowed from the analogous notion of {\it Lorentz frames} in Minkowski spacetime).
Note that all tensors constructed from the metric are unaffected by this transformation.
In addition,  $r_*$ preserves its functional form, $r_*=(\tilde v-\tilde u)/2$.

Consider two infalling observers, which move along two identical worldlines related to each other by a time translation. These observers cross the
event horizon (EH) at Eddington times $v_{eh}^1$ and $v_{eh}^2$, respectively (with
$v_{eh}^{(2)}>v_{eh}^{(1)}$).
Owing to this difference in $v_{eh}$, the two observers will not share the same function $v(\tau)$ (or $u(\tau)$).
To bridge this difference, we equip the second observer with its own Eddington frame
$(\tilde u, \tilde v)$, setting $\delta=v_{eh}^{(2)}-v_{eh}^{(1)}$ in Eq.  (\ref{EDshift}).
Since now $\tilde v_{eh}^{(2)}=v_{eh}^{(1)}$, it is not difficult to show that $\tilde v(\tau)$ of the second observer will be the same function as $v(\tau)$ of the first observer---and the same relation will apply between $\tilde u(\tau)$ and $u(\tau)$.

Consider now some linear perturbation field $\Psi(u,v)$ on the background spacetime (\ref{Eddington}). Like all other tensorial quantities, $\Psi$ is invariant under shifts in the Eddington frame.
We assume that near the CH $\Psi$ decouples to ingoing and outgoing components  (like all linear fields considered above); and we shall be concerned here with the field's outgoing component, which we denote $\Psi_u(u)$.

Pick two $u$ values $u_{1,2}$, with the only requirement that
$\Delta \Psi_u\equiv \Psi_2-\Psi_1 \neq 0$, where
$\Psi_{1,2}\equiv \Psi_u(u=u_{1,2})$.
We shall now examine how the second observer will experience this variation in
$\Psi_u$ from $\Psi_1$ to $\Psi_2$, as a function of its own proper time.
To this end we re-formulate the problem in terms of $\tilde u$ rather than $u$. The change from $\Psi_1$ to $\Psi_2$ thus occurs while the second observer moves from $\tilde u_1$ to $\tilde u_2$, where
\[
\tilde u_{1,2}\equiv u_{1,2}-\delta=[u_{1,2}-v_{eh}^{(1)}]+v_{eh}^{(2)} .
\]

Let us now fix $v_{eh}^{(1)}$, and yet consider the late-infall limit for the second observer: $v_{eh}^{(2)}\to \infty$.
Evidently, in this limit both $\tilde u_1$ and $\tilde u_2$ are pushed toward $\infty$.
The corresponding (second-observer) proper times $\tau(\tilde u_{1,2})$ will thus be pushed
to the same (finite) limiting value  $\tau(\tilde u\to\infty)$, that is, the moment of inner-horizon crossing.  In particular, the proper-time difference $\tau(\tilde u_2)-\tau(\tilde u_1)$
will vanish in this late-infall limit.

We conclude that at the late-infall limit, the finite variation $\Delta \Psi_u$ in the perturbation field $\Psi$ (which takes place between a certain pair of $u$ values $u_{1,2}$) occurs within a vanishing proper-time interval---at a moment which
(at the limit) coincides with that of inner-horizon crossing.
Thus we recover the effective shockwave phenomenon for late-infall observers.

\section{spherically symmetric non-linear perturbations}
\label{non-linear}

In this section we shift our focus from linear perturbations on a RN (or Kerr) background, to nonlinearly-perturbed BHs.
The main new ingredient is that now the infall orbit is disturbed by the MP, which in turn may influence the observer's experience of the perturbation.
For simplicity, we shall restrict attention here to a spherically-symmetric model of a nonlinearly-perturbed charged BH. We shall first present the model and describe the perturbed BH geometry, and then analyze the experience of late-infall observers in such a spacetime---demonstrating that the effective shock-wave phenomenon occurs in nonlinearly-perturbed BHs as well.

\subsection{Self-gravitating scalar field perturbations of a charged BH}
\label{non-linear_metric}

Let us consider a spherical charged BH perturbed by a spherically-symmetric self-gravitating scalar field. This model was investigated by several authors, primarily numerically \cite{GG,BS,Burko}  but also analytically
 \cite{ORIup, BOint}
(assisted by insights gained from earlier analytical investigations of the mass-inflation model \cite{PI,Ori91}).
The model consists of a massless, minimally-coupled, scalar field $\phi$, satisfying the covariant wave equation
$g^{\alpha\beta} \phi_{;\alpha\beta}=0$
on the (self consistently-perturbed)
metric of a spherically-symmetric charged BH.
The scalar-field energy-momentum tensor
\begin{equation}
T_{\alpha\beta}=\frac{1}{4\pi}\left(\phi_{; \alpha}\phi_{; \beta}
-\frac{1}{2}g_{\alpha\beta}\phi_{; \mu}\phi^{; \mu}\right)
\label{Tscalar}
\end{equation}
acts as a source term in the Einstein equations
(in addition to the electromagnetic contribution to energy-momentum), yielding a system of nonlinear field equations for the metric functions [e.g. $r(u,v)$ and $f(u,v)$ in Eq. (\ref{metric}) below].
As for initial conditions,
we consider here initial configurations wherein $\phi$ is initially compactly supported
outside the BH, as in \cite{Burko}.
(Alternatively, one may prescribe the initial data for $\phi$, including its presumed inverse-power tails, directly on the EH, as done in \cite{BS}.)
Evolving the initial data one then finds---not surprisingly---that at late times perturbations die out, and the BH settles down asymptotically to a member of the RN family, with charge $q$ and
a certain final mass $M$.
The scalar perturbations decay as
inverse-power tails. In particular, along the EH, $\phi\propto v^{-n}$ (typically with $n=3$) at late times \cite{Price, Bicak, Burko}.
These radiation tails fall into the BH and perturb its internal geometry.

The perturbed metric in the BH interior is conveniently expressed in double-null coordinates.
 In particular, in Eddington-like coordinates
\footnote{
The Eddington-like coordinate $v$
may naturally be defined in the perturbed spacetime by using characteristic initial-value formulation, and setting $r(v)$ and $f(v)$ along the outgoing initial ray (beyond the end of the compact initial support of $\phi$) to be the same functions as in the unperturbed RN spacetime
with the Eddington metric (\ref{Eddington})
(setting $F\to f$ therein).
The key property of $v$ is that it diverges at future null infinity, with $f\to 1$ on approaching the latter.
An analogous procedure may in principle be applied to define $u$,
yielding an ingoing null coordinate which diverges at the EH.}
we write the line element in the form
\begin{equation}
ds^{2}=-f(u,v)\, du\, dv+r^{2}(u,v)d\Omega_2^{2}.
\label{metric}
\end{equation}

The infalling scalar-field tail triggers the formation of a curvature singularity at the CH. This is a direct consequence of the infinite blue-shift that takes place at the inner horizon \cite{Penrose}, which leads to (almost) exponential divergence of the gradient of $\phi$---and of curvature.
 It turns out, however, that this is actually a {\it weak} \cite{Tipler, OriWeak}  curvature singularity, located exactly at the CH ($v\to\infty$).
 The metric tensor (in appropriate coordinates) approaches a finite, non-singular, limit as $v\to\infty$.
 For example,  we may use the Kruskal-like coordinates
\begin{equation}
U\equiv -e^{\kappa u}\, , \, \, \, V\equiv -e^{-\kappa v},
\label{Kruskal_in}
\end{equation}
with the line element
\begin{equation}
ds^2 = H(U,V)dUdV+ r^2(U,V) d\Omega_2^2
\label{Kruskal}
\end{equation}
(with $H= \kappa^{-2} f e^{v-u}$;
note that $U$ and $V$ are both future-directed, and correspondingly $H<0$).
The CH is located at $V=0$.
Both $r$ and $H$ have {\it finite, nonvanishing} values at the CH.
Yet, $\partial r/ \partial V$ diverges
at $V=0$, implying the presence of a null curvature singularity there (though a weak one).
The scalar field $\phi$ behaves in a manner similar to $r$: It is finite at $V\to 0$, yet $\partial \phi / \partial V$ diverges at that limit.

Perturbation theory predicts \cite{ORIup} (and numerical simulations \cite{Burko} confirm) that in the early portion (i.e. $u\gg M$) of the CH, the metric functions deviate only slightly from their respective values in the unperturbed RN solution. The domain $u,v\gg M$ is amendable to
perturbative treatment. Correspondingly we express $r$ and $H$ as
\begin{equation}
r(u,v)=r_{RN}(u,v)+\delta r(u,v)  \:,    \:\:   H(u,v)=H_{RN}(u,v)+\delta H(u,v)  \:,
\label{perturbed}
\end{equation}
where the suffix ``RN" denotes the corresponding function in the unperturbed RN spacetime. The perturbations $\delta r, \delta H$ vanish in the limit $u,v\to \infty$.
(This limit corresponds to the past boundary of the CH, but also to $t \to \infty$ along spatial lines of constant $v-u$.)
In the domain $u,v\gg M$, the scalar field is dominated by its linear perturbation term, and the MP $\delta r,\delta H$ by the second-order perturbation, as described below
(see Appendix).

As was mentioned above, $r$ (like $H$ and $\phi$) is finite along the CH. It initially starts at $u\to \infty$ with $r=r_{RN}=r_-$, but subsequently shrinks steadily with time ($-u$), due to the focussing induced by the outflux of scalar-field energy-momentum across the CH. At some stage $r$ shrinks to zero---at which point the null weak CH singularity terminates, and connects to a strong, spacelike, $r=0$ singularity \cite{GG, BS, Burko}.

\subsection{Late-infall orbits}
\label{nonlinear_orbits}

We turn now to investigate the experience of late-infall observers in this spacetime of a nonlinearly-perturbed spherical charged BH.
We consider (equatorial) infalling geodesics which are not necessarily radial.
The angular-momentum parameter $L$ is conserved in these geodesics, though $E$ is no longer conserved.

For concreteness we shall focus here on the behavior of the
metric function $r$, which the infalling observer probes as a function of his proper time $\tau$.
Physically, a rapid change in $r$ will imply (for a finite-size observer) a rapid deformation in the tangential directions $\theta,\varphi$.
 Again, we choose two $u=const$ hypersurfaces, denoted $u=u_{1,2}$ (with $u_1>u_2$), requiring that both hypersurfaces intersect the CH
 ({\it before} $r$ shrinks to zero).
While the observer progresses from $u=u_{1}$ to $u=u_{2}$,
$r$ changes by the amount
$\Delta r \equiv r_2-r_1 $.
Hereafter a quantity with a sub-index ``1" or ``2" will denote the value of this quantity as the worldline crosses the hypersurface $u_1$ or $u_2$, respectively.
Since along the CH $r$ is steadily shrinking,
one finds that $\Delta r \neq 0$.

Let us now evaluate the proper-time interval $\Delta \tau \equiv \tau_2 - \tau_1$,
using
\[
d\tau=\left(-g_{\alpha\beta}\frac{dx^\alpha}{dU}\frac{dx^\beta}{dU}\right)^{1/2} dU
=\left[ |H| \frac{dV}{dU}-r^2 \left(\frac{d\varphi}{dU}  \right)^2 \right]^{1/2} dU \, .
\]
Note that $H$ (like $r$) is
bounded throughout the domain $u_1>u>u_2$,
so we can easily bound $\Delta \tau$ by
\[
\Delta \tau < (H_{max})^{1/2} \int_{U_1}^{U_2}
  \left( \, \frac{dV}{dU}\right)^{1/2} dU \, ,
\]
where $H_{max}$ denotes the maximal value of $|H|$ in the worldline's section between $u_1$ and $u_2$.
The integral on the RHS is nothing but the proper-time length of a timelike curve connecting the two points $(U_1,V_1)$ and $(U_2,V_2)$, in a fiducial two-dimensional spacetime with the flat metric $ds^2= - dU dV$. It is bounded above by the (timelike) geodesic connecting these edge points, whose length is
$ \left( \Delta V \, \Delta U \right)^{1/2}$, where $\Delta U=U_2-U_1$ and
$\Delta V\equiv V_2-V_1=e^{-\kappa v_1}-e^{-\kappa v_2} $.
Clearly, $\Delta V<e^{-\kappa v_1}< e^{-\kappa v_{eh}}$, therefore
\[
\Delta \tau < (H_{max}\Delta U)^{1/2}  \, e^{-\kappa v_{eh} / 2} .
\]

Consider now the late-infall limit, which is the limit of large $v_{eh}$.
In this limit $V_{1,2}\to 0$. Therefore $H_{max}$ approaches $H_{max}^{ch}$, which is the maximal value of $|H|$ along the section $u_1>u>u_2$ of the CH. We obtain our bound on $\Delta \tau$ (for late-infall observers) in its final form:
\begin{equation}
\Delta \tau < B e^{-\kappa v_{eh}/2} ,
\label{tauBound}
\end{equation}
where $B\equiv ( H_{max}^{ch} \Delta U)^{1/2} $ is a parameter which depends on $u_1$ and $u_2$ but not on the orbit's infall time.

We conclude that late-infall observers will measure a non-vanishing variation $\Delta r$ in the metric function $r$, within a short proper-time difference $\Delta \tau$ which shrinks exponentially in the infall time $v_{eh}$---which is again an effective shock-wave phenomenon.

\subsection{Do the late-infall orbits cross the CH?}
\label{NOcrossing}

Our analysis so far did not
make use of the perturbative nature of the metric field (at the early portion of the CH).
We merely assumed that the CH singularity is null and weak---and more specifically, that $H$ admits a finite limiting value along the CH.
Correspondingly, there was no need to restrict $u_1$ and $u_2$ to the perturbative domain ($u\gg M$):
We only required that the surfaces $u=u_1$ and $u= u_2$ intersect the CH,
rather than the spacelike singularity.
However, there still was one hidden assumption: We implicitly assumed that the late-infall observers will make it all the way from $u_1$ to $u_2$ {\it without intersecting the CH} (that is, with finite $v$). Once an observer intersects the CH, we cannot make any concrete statement about his subsequent experience, because the CH is by definition the boundary of the domain of unique prediction (for e.g. the metric functions).
\footnote{Furthermore, owing to the divergence of curvature at the CH singularity, it is unclear whether a classical extension beyond the CH will be physically meaningful. }

We therefore still need to complete this missing piece of the analysis.
We shall show that as long as $u_2$ is located in the weakly-perturbed domain of the CH (that is, $u_2$ is large compared to $M$), the late-infall orbits indeed arrive at $u=u_2$ with finite $v$.

The control on the growth of $v$ will be achieved by monitoring the evolution of the geodesic's ``energy" parameter
\begin{equation}
E\equiv -(u_{u}+u_{v}).
\label{Edefined}
\end{equation}
Note that $E$ (unlike $L$) is no longer conserved, because the perturbations destroy the exact $t$-translation invariance of the RN background. Yet, following the evolution of $E$ will enable us to control $u^v$, and thereby the evolution of $v$ along the orbit.

\subsubsection{Equation of motion for $E$}

The lower-index covariant geodesic equation, applied to the Eddington-like metric (\ref{metric}), reads
\[
\dot{u}_{\mu}=(1/2)g_{\alpha\beta,\mu}u^{\alpha}u^{\beta}=-(1/2)f_{,\mu}u^{u}u^{v}+(L^{2}/r^{3})r_{,\mu} \, .
\]
Therefore
\[
\dot{E}=-(\dot u_{u}+\dot u_{v})=(1/2)(f_{,u}+f_{,v})u^{u}u^{v}-(L^{2}/r^{3})(r_{,u}+r_{,v}) \, .
\]
To get rid of the term $u^{u}u^{v}$ in the RHS, we use the normalization condition
$g_{\alpha\beta}u^{\alpha}u^{\beta}=-1$,
which for the metric (\ref{metric}) yields
\begin{equation}
u^{u}u^{v}=\left(1+L^{2}/r^{2}\right)/f.
\label{eq:normalization}
\end{equation}
The quantity $(f_{,u}+f_{,v})u^{u}u^{v}$ then becomes
\[
\left(1+L^{2}/r^{2}\right)\left[(\ln |f|)_{,u}+(\ln |f|)_{,v}\right].
\]
Noting that
the last term in squared brackets is equal to
$(\ln |H|)_{,u}+(\ln |H|)_{,v}$,
we re-express $\dot E$ in terms of the Kruskal-like metric function $H$:
 \begin{equation}
\dot{E}=\frac{1}{2}\left(1+\frac{L^{2}}{r^{2}}\right)\left[(\ln |H|)_{,u}+(\ln |H|)_{,v}\right]
-\frac{L^{2}}{r^{3}} (r_{,u}+r_{,v}).
\label{eq:evolvE}
\end{equation}

Note that in the background RN metric,
both $H$ and $r$ are functions of $v-u$, hence $E$ is conserved.
Evolution of $E$ will thus only result from the MP
$\delta r$ and $\delta H$.

So far all equations were exact.
To proceed beyond this point, we restrict attention to the early portion of the CH, where perturbations are presumably small,
and carry out the analysis at the leading order in the MP.
\footnote{
That is, we expand the various prescribed {\it background functions} (i.e. the functions of $u,v$ obtained from the metric functions etc.) to first order in the MP. However, we do {\it not} linearize the worldline-related quantities, like $E$, $\dot E$, $u^v$, $u^u$, etc.}
Also, since we are dealing with the orbit's evolution very close to the CH,
we may replace the background's functions $r_{RN}$ and $H_{RN}$ by their corresponding inner-horizon values $r_-$ and $H_-\equiv H_{RN}(r=r_-)$.
We obtain
\begin{equation}
\dot{E}=A(\delta H_{,u}+\delta H_{,v})+B(\delta r_{,u}+\delta r_{,v})
\equiv W(u,v),
\label{eq:Edot}
\end{equation}
where $A$ and $B$ are constants,
\[
A=\frac{1}{2H_-}\left(1+\frac{L^2}{r_-^2}\right) \,, \,\,\,B=-\frac{L^2}{r_-^3} \, .
\]
Recall that $H_-$ (like $r_-$) is a finite, non-vanishing constant.

\subsubsection{Analyzing the evolution of $E$ }

We re-write Eq. (\ref{eq:Edot}) as
\begin{equation}
dE/du=W(u,v)/u^{u},
\label{eq:dEdu}
\end{equation}
in which we view $u^u$ as a (yet unknown) function of the parameter $u$ along the geodesic. One might choose to approximate this function $u^u(u)$ by the corresponding function for geodesics in the
unperturbed RN geometry.
We shall not proceed here in this way,
because we do not want to assume a-priori that the accumulating perturbations in the orbital parameters must be small.
Instead, we shall proceed by expressing $u^{u}$ in terms of $E$.
To this end we use the contravariant version of Eq. (\ref{Edefined}),
namely
\[
u^u+u^v=2E/f .
\]
This, together with Eq. (\ref{eq:normalization}), constitutes
a closed algebraic system for the two unknowns $u^{u},u^{v}$.
One can of course write down the exact solution of this algebraic system. However, it will be more instructive to employ here the approximate solution, associated with the smallness of $f$:
We are dealing here
with the near-$r_{-}$ region, where
\footnote{Note that $\delta H\ll H_{RN}$ (valid in the perturbative domain considered here) also implies that $\delta f$ (the perturbation in $f$) satisfies $\delta f\ll f_{RN}$; that is, $f\cong F \propto e^{\kappa (u-v)}$.}
$f \propto e^{\kappa (u-v)} \ll1$.
The algebraic system thus yields the simple approximate solution (to leading order in the small parameter $f$)
\footnote{The algebraic system also admits a second solution, in which
$u^u$ and $u^{v}$ are interchanged. However, since our late-infall observers
enter the $r\approx r_{-}$ region with large $u^{u}$, it is
the solution (\ref{eq:uUPvalues})
which actually takes place.}
\begin{equation}
u^{u}\cong 2 E/f,\quad u^{v}\cong \left(1+L^{2}/r^{2}\right)/(2E).\label{eq:uUPvalues}\end{equation}
Substituting this expression for $u^u$ in Eq. (\ref{eq:dEdu}) we find
\begin{equation}
d(E^{2})/du=W f = \left[ A(\delta H_{,u}+\delta H_{,v})+B(\delta r_{,u}+\delta r_{,v}) \right] f .
\label{eq:dE2}
\end{equation}
Setting $f\cong F\cong const \times e^{\kappa (u-v)} $ (where, recall, $F\equiv f_{RN}$), we obtain
\begin{equation}
\frac{d(E^{2})}{du}
= \left[ \tilde A(\delta H_{,u}+\delta H_{,v})+\tilde B(\delta r_{,u}+\delta r_{,v}) \right] e^{\kappa (u-v)} ,
\label{eq:dE2f}
\end{equation}
where we have absorbed
the above $const$ (in $F$)
in $\tilde A$ and $\tilde B$. Note that the RHS in this equation is a prescribed function of $u$ and $v$
(with no reference whatsoever to four-velocity).

At this point it will be useful to refer to the concrete form of the MP $\delta r$ and $\delta H$. We focus here on the early portion of the CH (i.e. $u\gg M$), where the MP
are small and decay as inverse powers of $u$ and/or $v$.
We denote by $\Delta E$ the modification in $E$ acquired in the near-CH region, up to some $u_2\gg M$, due to the presence of MP.
Based on the inverse-power form of the MP, in the Appendix we derive the bound
\begin{equation}
 \left| \Delta E  \right|  <
C \, (v_{eh})^{-(2n+1)} =C \, (v_{eh})^{-7} ,
\label{DeltaEBOUNDfinal}
\end{equation}
where $C$ is a certain parameter (independent of $v_{eh}$).
In particular, we find that for late-infall orbits $|\Delta E|\ll E$.

\subsubsection{Analyzing the evolution of $v$ }

We proceed now to analyze the evolution of $v$, showing that it remains finite throughout $u\geq u_2$ (for any $u_2 \gg M)$. From Eq. (\ref{eq:uUPvalues}) we have
\begin{equation}
\frac{dv}{du}=\frac{u^{v}}{u^{u}}
\cong  \frac{f}{4E^2} \left(1+L^{2}/r^{2}\right) .
\label{uv}
\end{equation}
Now for late-infall orbits we already found that $|\Delta E|\ll E$ and hence we may regard $E$ as constant (essentially the entrance value of $E$). Also we may set $r\cong r_-$, and take the near-CH form of $f$, namely
$f \cong -e^{\kappa(u-v)} \times const $.
We obtain
\begin{equation}
\frac{dv}{du}
\cong - \tilde{C} \,  e^{\kappa(u-v)} ,
\label{uva}
\end{equation}
where $\tilde{C}$ is some positive constant.
Re-writing this as $d(e^{\kappa v})/d(e^{\kappa u}) \cong  - \tilde{C} $,
we obtain
\begin{equation}
e^{\kappa v} \cong - \tilde{C} e^{\kappa u} + const \,  .
\label{dUV}
\end{equation}
[Notice that the last two equations are the same as those describing late-infall geodesics in the {\it unperturbed} RN geometry.]
This expression is bounded above (by the $const$ in the RHS).
Thus, $v$ remains finite throughout the range $u\gg M$---meaning that the orbit cannot cross the CH (located at $v\to \infty$) in that domain.

\subsection{Concluding Remarks}

We found that the late-infall observers cannot cross the CH in the regime $u\gg M$.
The analysis in section \ref{nonlinear_orbits} then shows that the proper-time for these observers to move from $u=u_1$ to $u=u_2$, for any  $u_1>u_2\gg M$, decreases exponentially in the infall time $v_{eh}$.
Since $r$ varies during this range by a finite amount $\Delta r<0$, we inevitably face here the phenomenon of effective {\it gravitational shock-wave}: A discontinuity in the metric tensor, which propagates along a null hypersurface (the outgoing section of the inner horizon in our case). Physically, this means that any extended object will undergo a sudden deformation, by a
certain amount, within an effectively-vanishing proper time.

The amplitude of such a gravitational shock-wave may naturally be characterized by the (dimensionless) magnitude of the object's deformation. Specifically in the spherical model studied here, the shock's amplitude may be taken to be the dimensionless quantity
$|\Delta r|/r_-$.

Two additional remarks are in order here:

\begin{enumerate}
\item Since the discussion in section \ref{NOcrossing} assumed $u_2\gg M$ in order to treat perturbations perturbatively, one might mistakenly conclude that the amplitude of the gravitational shock wave can be weak (i.e. $|\Delta r|/r_- \ll 1$).   But this is not the case.  It is clear from the above analysis that late-infall geodesics cannot fall across the CH until after the perturbations grow to be of order 1.  We therefore see from section \ref{nonlinear_orbits} that late freely-falling observers must face a strong gravitational shock wave
whose amplitude is at least of order 1.

\item
In order to allow a simple discussion of the quantity $E = -(u_u + u_v)$, which is conserved along geodesics in exact RN, we have so far assumed freely-falling worldlines.  But let us now consider an accelerated late infall-time observer.  We choose the acceleration as a function of proper time to be bounded and such that, in the unperturbed RN geometry, the worldline would reach the outgoing inner horizon before crossing the ingoing inner horizon.  Note that in the limit of large $v_{eh}$ all of these accelerations occur before reaching $u=u_1$.  As a result, in the unperturbed spacetime $E$ becomes some constant $E_{final}$ for all $u  < u_1$.  Corresponding late infall-time observers in the perturbed spacetime may thus be analyzed just as for the freely-falling observers discussed above but with the entrance value of $E$ replaced by $E_{final}$.  We conclude that any accelerated late infall-time observer  who ``would have reached the outgoing inner horizon in unperturbed RN"  also experiences a shockwave in the perturbed spacetime.
\end{enumerate}

\section{Effective spacetimes for late-infall observers}

\label{lateT}

We have seen above that, in the limit of late infall times, observers who enter a perturbed Reissner-Nordstr\"om black hole experience an effectively unperturbed Reissner-Nordstr\"om geometry up to the point where they would expect to encounter an inner horizon at $r=r_-$.  At this point, observers who would have reached the outgoing inner horizon in exact RN then encounter a shockwave across which the metric changes discontinuously.  

Describing the detailed nature of this discontinuity requires further investigation.  While we will not attempt a precise treatment here, it is natural to expect that the above shockwave in fact contains a curvature singularity, as we know that our observers will eventually reach such a singularity and we expect that, since they are already ``nearly null'' in the region described in section \ref{non-linear}, all proper times along their worldlines will be compressed to zero in the large $v_{eh}$ limit\footnote{\label{othmat} Note that this argument is equally valid if, for some kinds of matter, the spacetimes contain no spacelike singularity.  In such cases the ingoing worldline will reach the weak null singularity, which is also a curvature singularity.}.

In addition, it is clear that some accelerated observers will reach the ingoing weak null singularity shown in figure \ref{summary} (left).  For observers who enter the black hole very late, in the region where perturbations are very small, this may be accomplished using roughly the same set of accelerations (as defined in their own reference frame) as would be required to reach the ingoing part of the inner horizon in an unperturbed RN black hole.  Furthermore, in the limit of late infall times, such observers arrive at the portion of the weak null singularity close to $i^+$ in figure \ref{summary} where the singularity is extremely weak, so that again such observers will measure no noticeable departure from unperturbed RN until $r$ is exceedingly close to $r_-$.

To compliment the precise calculations of sections \ref{linear} and \ref{non-linear}, we find it useful to give a brief  heuristic argument that reinforces the above statements.  To do so, note that taking our observer to enter the black hole at late times is equivalent to considering some fixed infall time and replacing the given spacetime with one in which the black hole formed (along with its full set of perturbations) at a much earlier time.  As a particular example, we might consider a scenario which starts with empty Minkowski space in the interior and where both the matter that forms the black hole and our observer are dropped in from some large value of $r$ at finite advanced times $v_{matter}$ and $v_{obs}$ respectively.  For example, the observer might be ejected from a static space station.  The matter could be dropped from a dense network of such space stations, or simply assumed to cross the $r$-coordiante associated with the observer's space station near the advanced time $v_{matter}$. The observer is also given some particular instructions for executing accelerations as a function of proper time along his worldline and we take the matter to be released as some particular function of advanced time peaked around some $v_{matter}$.  Allowing $v_{matter}$ to vary leads to a one-parameter family of (isometric) spacetimes.  We wish to study the limit $v_{matter} \rightarrow -\infty$ holding $v_{obs}$ fixed.

It is useful to assume that the perturbed spacetime preserves spherical symmetry.  The advanced time $v$ is thus a well-defined null coordinate everywhere in the globally hyperbolic part of the spacetime (shown at left in figure \ref{summary}) and the retarded time $u$ is a similarly well-defined null coordinate in the region outside the event horizon.  Furthermore, it is clear that two spacetimes differing by a shift of $v_{matter}$ can be described by essentially the same metric functions, albeit again with a corresponding shift of $u,v$.  For example, introducing $\tilde u = u-v_{matter}$ and $\tilde v = v - v_{matter}$, we may take the collapsing spacetimes to be described by the metric

\begin{equation}
\label{uvform}
ds^2 = - f_{col}(\tilde u,\tilde v) \ du \ dv + r_{col}^2(\tilde u,\tilde v) d\Omega_2^2
\end{equation}
for some fixed functions
$f_{col}(\tilde u,\tilde v)$, $r_{col}(\tilde u,\tilde v)$
which do not explicitly depend on $v_{matter}$.
Similarly, the center of spherical symmetry is described by some fixed curve $\tilde u = c_{col}(\tilde v)$ and the event horizon is the curve
$\tilde u = + \infty$.  The fact that for any fixed $v_{matter}$ the spacetime approaches RN as $u,v \rightarrow \infty$ implies that in the formal $v_{matter} \rightarrow -\infty$ limit with $u,v$ fixed we have
$r_{col}(u-v_{matter},v-v_{matter}) \rightarrow r(u,v)$ and $f_{col}(u-v_{matter},v-v_{matter}) \rightarrow F(r(u,v))$, where $F(r)$ and $r(u,v)$ are the functions defined in section \ref{preliminaries} for exact RN.

We would like to find a similar construction for the region to the future of the event horizon.  In that region the retarded time $u$ cannot be defined simply by tracing outgoing radial null rays to $I^+$.  Rather than attempt to find a physically preferred definition of $u$, let us therefore consider any null coordinate $u$ which is smooth inside the event horizon, has past-directed gradient $\nabla_a u$,  and for which the metric
again takes the form (\ref{uvform}) on each member of our family of spacetimes where the functions $f_{col},r_{col}$ again approach $F(r(u,v))$ and $r(u,v)$ as defined in section \ref{preliminaries} in the limit of large $u,v$ with fixed $v_{matter}$.  Note that inside the event horizon $\tilde u$ will decrease to some $v_{matter}$-independent minimum value $u_{min}$ at the point where the center of symmetry hits the singularity.  We choose this $u_{min}$ to be finite.

\begin{figure}[ht]
\begin{center}
\includegraphics[scale=.5]{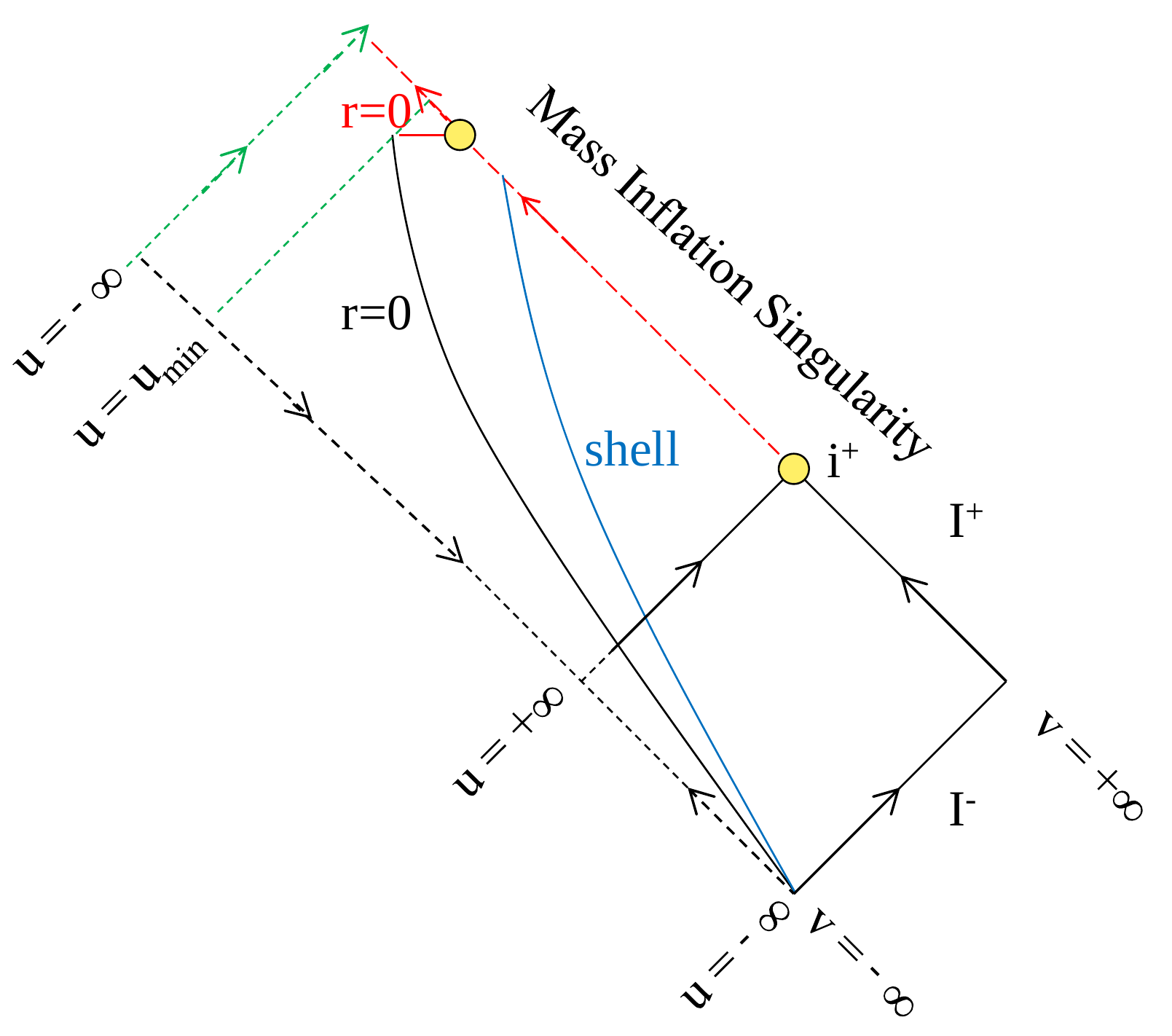}
\includegraphics[scale=.5]{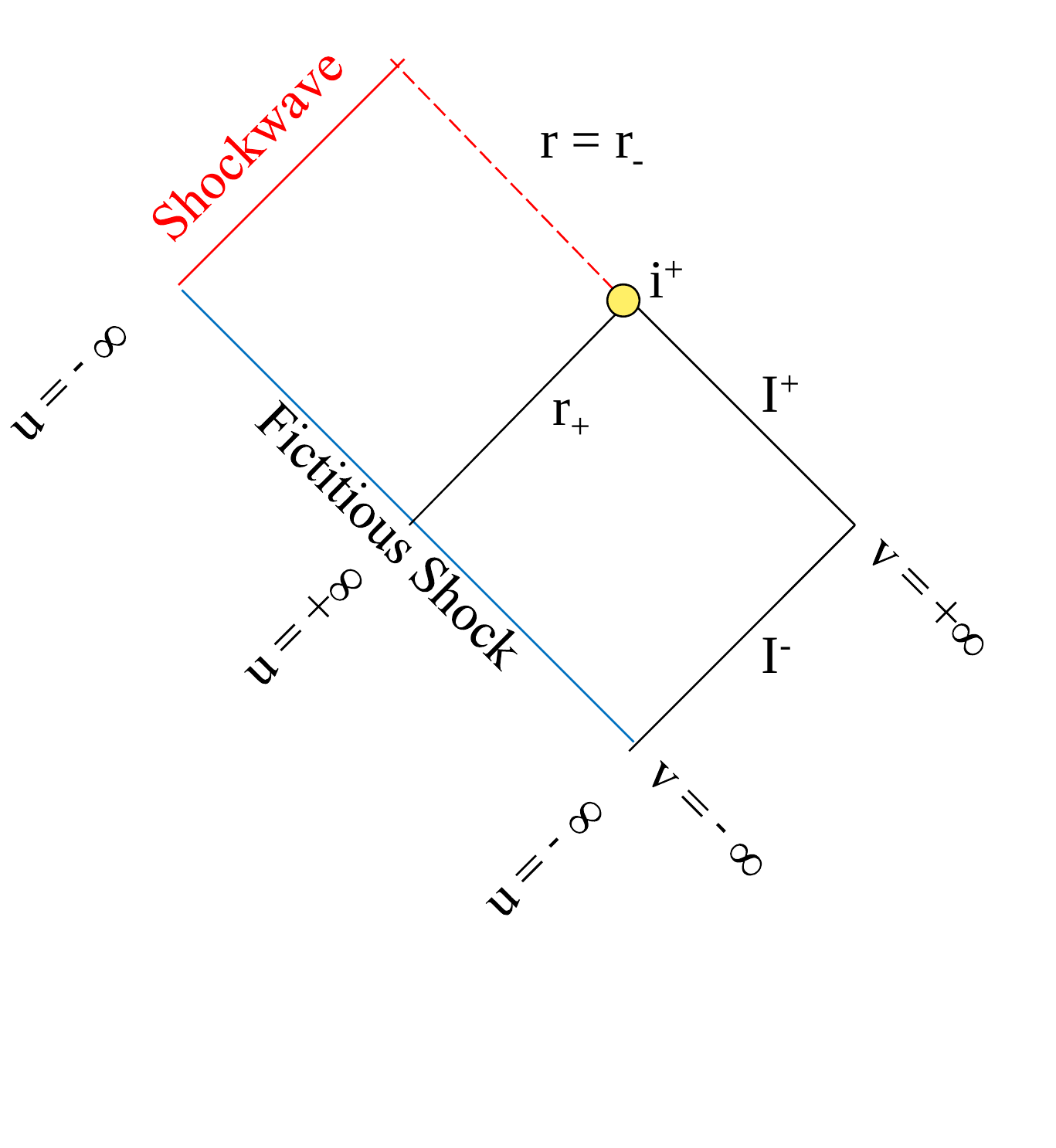}
\caption{\label{pertT}
 {\bf Left:} Some lines of constant $u,v$ drawn on the spherical spacetime describing a collapsing shell perturbed by a massless scalar field.  The long-dashed line (top right boundary) is the mass-inflation singularity.   In contrast, the short dashes mark coordinate lines lying outside the physical spacetime (since $r < 0$).
The vector field $\partial_u + \partial_v$ is tangent to any surface $u = \pm \infty$ or $v = \pm \infty$.  The arrows indicate the direction of this vector field along such surfaces.  The point marked $i^+$ is fixed under the action of this vector field.
{\bf Right:}
The effective geometry for late-time observers obtained by flowing the interior of our spacetime {\it backward} along the vector field $\partial_u + \partial_v$, while similarly deforming the boundaries. As for figure 1 (right), it consists of the region of the unperturbed eternal Reissner-Nordstr\"om black hole with $r > r_-$, together with certain shockwave singularities.  The shock along the lower left boundary descirbes a jump from $r=r_+$ to $r=0$ (corresponding to the regular center of spherical symmetry shown in the left diagram).
Though it arises only in the formal late-infall limit, this shock is not actually accessible to any late-infall observer.  In this sense the shock is fictitious.  In contrast, the shock along the upper left boundary is accessible to late-time observers.  It contains  the final piece of the shell worldline and describes a jump from $r=r_-$ to $r=0$ (corresponding to the spacelike singularity shown in the left diagram).
}
\end{center}
\end{figure}

We expect that late-time observers experience an effective spacetime described by the formal
$v_{matter} \rightarrow -\infty$
limit (with $u,v$ held fixed) of (\ref{uvform}) for {\it some} definition of the internal $u$ coordinate above\footnote{For example, if the metric coefficients of such a collapsing spacetime can be analytic in some null coordinates, then the action on the interior metric due to a shift of $v_{matter}$ will be determined by analytic extension of the action on the exterior metric, and this interior action can be used to define an interior null coordinate $u$ that simply shifts with $v_{matter}$.  It is clear that the resulting $u$ will have the properties required above near $i^+$.}.  Moreover, the essential features of this limit are independent of the particular definition of $u$. Indeed, by construction this limit is described by $f_{col} \rightarrow F(r(u,v))$ and $r_{col} \rightarrow r(u,v)$ on the domain $u > u_{min} + v_{matter} \rightarrow - \infty$.   All of the additional structure has been sent off to $u = -\infty$, which of course nevertheless lies at finite affine parameter along future-directed geodesics.  Thus, as shown in figure \ref{pertT}, this limiting spacetime is just the region of unperturbed RN to the future of its past horizon and to the past of its outgoing inner horizon.  However, rather than being smooth, the outgoing null surface $u = -\infty$ is a shockwave into which all of the time-dependent structure of the spacetime has been compressed.  We should also recall that, although it is no longer apparent from the limiting form of the metric, the entire ingoing surface $v = +\infty$ to the future of $i^+$ is a curvature singularity (i.e. the mass-inflation singularity)
 for any finite value of $v_{matter}$.

\section{Discussion}
\label{disc}

Our work above examined the experiences of observers who enter classical spherically-symmetric charged asymptotically flat black holes subject to both linear and non-linear perturbations.  We also considered linear perturbations of Kerr black holes.   Our emphasis was on the limit of late infall times, $v_{eh} \rightarrow \infty$.  In this limit, our observers' observations
agree precisely with those of similar observers in the unperturbed (stationary) black hole spacetime up to the point where the latter observers would reach the inner horizon.    At that point, however, those late infall-time observers who would have reached the (smooth) outgoing inner horizon in
unperturbed RN
instead encounter an effective gravitational shock-wave in the perturbed spacetime. The shock's width (expressed in terms of the proper time of infalling observers) decreases with infall time as $e^{-\kappa v_{eh}}$,  hence it effectively vanishes for late-infall observers.

This may be thought of fundamentally a time-dilation effect.  Recall that the Killing field $\partial_t$ of a stationary black hole acts like a Minkowski-space boost in the region near where the ingoing inner horizon meets the outgoing inner horizon.  This is precisely the region in which late-infall observers encounter any structure present in the black hole interior.  As a result, observers who enter at later and later times arrive here in more and more highly boosted reference frames.  Thus they approach at nearly the speed of light and transit through in vanishing proper time, experiencing any structure as a shock-wave.

In such a gravitational shock-wave, the metric undergoes a discontinuity and changes by a finite amount
\footnote{The magnitude of the shock-wave may also be infinite in some cases. In the discussion here we shall assume it is finite.}
in vanishing time. An extended physical object hitting this shockwave will thus undergo a sudden deformation (i.e. shear and/or contraction), by a finite amount, within an effectively-vanishing proper time. Owing to their short time scale, the gravitational tidal forces entailed in the shock will presumably dominate over all
internal interactions. Thus, each single nucleon (say) will be deformed by a certain amount (typically of order 1) while traversing such a shock-wave.

We also gave a heuristic argument that the experiences of late-time observers are described by the simple effective spacetime shown at right in figure \ref{pertT}, which in particular reproduces the above exact results.   On the basis of this argument one expects the outgoing shockwave to contain a strong curvature singularity\footnote{For the case where the perturbations are associated with a massless scalar field. For more general matter fields, the shockwave may contain only a weak curvature singularity. See footnote \ref{othmat}.}, with the area-radius $r$ shrinking to zero across the shockwave for the case described in figure \ref{pertT}.  However, these final details remain to be verified by other techniques.

While the explicit non-linear analysis in section \ref{non-linear} was restricted to the spherically-symmetric case, we strongly expect that the same phenomenon of shockwave formation will take place in nonspherical black holes as well, and particularly in perturbed spinning black holes. This expectation is based on the combination of several pieces of evidence:
the experience of late-infall observers inside a linearly-perturbed Kerr BH (discussed in section \ref{linear}),
the sub-dominance of nonlinear perturbations in generically-perturbed spinning black holes (particularly near the early portion of the CH),
and---more generally---on the profound similarity between the inner structures of charged and spinning black holes.

It may be interesting to compare the properties of the two different types of null singularity that develop at the inner horizon: the shock-wave singularity at the latter's outgoing section, and the weak curvature singularity at the CH. We argue that from the physical point of view the shock-wave singularity is the more violent one---particularly for very late-infalling observers.  To see this, consider two representative orbits in the RN geometry: Orbit (a) hits the ingoing section of $r=r_-$, whereas orbit (b) hits the latter's outgoing section. Let us now consider the experiences of these two observers on their approach to $r=r_-$, in case the RN spacetime is perturbed---focussing our attention on the late-infall orbits belonging to the two families (a,b). Observers of type (a) will hit a true curvature singularity at the CH. This singularity is weak, however, and any observer will only experience a finite integrated deformation up to the CH. Furthermore, this deformation will decrease with increasing $v_{eh}$ (as an inverse power thereof), and will vanish
in the late-infall limit.
 On the other hand, observers of type (b) will experience, on approaching $r=r_-$, a certain amount of deformation
which does {\it not} decrease with increasing $v_{eh}$.
This is because the shock's amplitude approaches a certain limiting value at large $v_{eh}$
(it is only the shock's width which evolves at late time, it decreases exponentially in $v_{eh}$).
Thus, at least as far as the overall tidal deformation is considered, at the late-infall limit observers of type (a) will effectively feel no singularity at the CH, whereas late-infall observers of type (b) will feel a violent gravitational shock-wave singularity, whose magnitude is typically of order unity (at least).

While our considerations were completely classical, our conclusions also suggest properties of quantum black hole microstates.  In particular, the second law of thermodynamics implies that the late-time limit of any black hole spacetime should approximate the typical quantum state of the relevant ensemble.  It is thus natural to conjecture that the typical state in an ensemble of charged or rotating black holes corresponds to the geometry displayed in figure  \ref{pertT} (right) together with quantum corrections, as opposed to say that shown in \ref{pertT} (left).

Some of our results carry over directly to generic (stable) black holes in any dimension.  Indeed, in retrospect our linear analysis depended only on the general form of a (future-directed) time-translation $\partial_t$ between the inner and outer horizons of the unperturbed black hole, and in particular on the pattern of surfaces invariant under its action.   The essential ingredients are just that: i) the (smooth) inner and outer horizons are invariant under $\partial_t$, ii) the point $i^+_{right}$ in figure \ref{RNfig} (left) is an attractive fixed point of $\partial_t$ while $i^+_{left}$ is a repulsive fixed point, and iii) perturbations in the immediate vicinity of the outgoing inner-horizon decouple into independent $u$ and $v$ components\footnote{
At least for scalar fields with the usual kinetic term, one may derive (iii) from (i) and (ii) by defining $\tilde V = \lambda V$ and taking the limit $\lambda \rightarrow \infty$ to zoom in on the region near $V=0$. Corresponding results may also hold for higher spin fields.}.   Whenever these features arise, late infall-time observers who reach the outgoing inner horizon will experience any given linear perturbation as a shockwave.  Note that, despite our use of Eddington coordinates $u,v$ in the main text, this more geometric form of the argument is manifestly coordinate invariant.

At the non-linear level, the assumption that the perturbed spacetime approaches a stationary geometry at $i^+_{right}$ which is an attractive fixed point of $\partial_t$ again implies that late infall-time observers experience an essentially unperturbed solution up to the point where they would expect to reach an inner horizon.  One is tempted to draw a conformal diagram similar to that in figure \ref{pertT} (right) and again conjecture that infalling observers experience an effective outgoing shockwave.  However, investigating this conjecture in detail will require more sophisticated methods or, perhaps, numerical simulations.

Throughout this work we have considered what one may call `test observers,' which experience perturbations of the spacetime but do not source further perturbations.  In contrast, any physical observer who falls into the black hole may be expected to create additional perturbations which now typically fall into the black hole at an advanced time comparable to $v_{eh}$ (and thus which cannot generally be ignored simply by taking the limit of large $v_{eh}$).  It would be interesting to understand what this implies for the experiences of such physical observers.  For example, consider an observer who reaches the would-be ingoing inner horizon (the CH).  Due to the decay of other perturbations as reviewed in section \ref{non-linear_metric}, for
sufficiently large $v_{eh}$ the perturbations at the relevant part of the CH will generically be dominated by those sourced by our observer himself.
One would expect him to encounter a null weak singularity whose strength (typically tiny, of order the mass ratio between the BH and observer) is determined by perturbations of his own creation.
In contrast, even if our observer instead reaches the would-be outgoing inner horizon, we see no reason for perturbations sourced by the observer to destroy the outgoing shock wave set up by earlier perturbations or to significantly change the experience described above.

In our work above, we assumed that the spacetime approaches some non-extreme stationary black hole geometry near $i^+$.  To simplify the analysis, we assumed ``generic'' parameter values for this limiting BH, and in particular that it is not too close to extremality.  While extending these results to nearly-extreme black holes is not difficult (see in particular footnote \ref{extend} in appendix A), it remains to understand the precisely extreme case.  Taking the formal extreme limit of our late-time results (i.e., first taking $v_{eh} \rightarrow \infty$ with $r_+ \neq r_-$ and then taking $r_- \rightarrow r_+$) agrees with the picture suggested in \cite{extremes} (and supported by \cite{garf}) in which late-time observers experience a singularity which effectively resides at the final event horizon.
This suggests that such a picture should hold
 at least for horizons that approach extremality sufficiently slowly relative to the production of perturbations (and relative to the time scale set by the surface gravity).
 This in particular should apply for large black holes when the approach to extremality is due to quantum processes in cases where the extreme BH is quantum mechanically stable (e.g., with enough supersymmetry).  However, a complete analysis is again left for future work.

To simplify the discussion, our analysis has focussed on very late time observers who enter asymptotically flat black holes.  But let us now comment on the implications for observers who enter astrophysical black holes at finite times.  So long as the current accretion rate and any effects from the expansion of the universe are small  (e.g, at the 1\% level), and so long as a few light-crossing times have passed since the black hole experienced a large perturbation (including the initial formation of the black hole), a finite infall-time observer who reaches the would-be outgoing inner horizon should also experience our effective shock wave. This can be seen from the following points: (i) By causality, nothing that enters the black hole more than a few light-crossing times to the future of our observer can affect his experience.   Thus the experience of our observer can be described by modeling the astrophysical black hole as an asymptotically flat black hole with a small accretion rate.  (ii)  Although in various parts of this paper we have focused on the internal structure of the BH at very late time (after the decay of radiative tails), one should bear in mind that the process of shock formation is in fact rather quick: Owing to the exponential decrease of the shock width, a $v_{eh}$ value of only a few light-crossing times is required for the very narrow shock configuration to build up. During such a $v_{eh}$ interval of, say, 10-20 times $M$, a weak accretion will not have a chance to cause a significant modification to the process of shock formation.

For observers who fall into such an accreting BH with a much larger $v_{eh}$, the accumulating effects of continuing accretion may possibly be more significant (see e.g. \cite{accreting}). Nevertheless, in the limit of weak accretion, any effects of this accretion can cause only small changes from the scenario described above.  While this might give our effective shock a finite width (independent of $v_{eh}$) or perhaps make the effective shock timelike or spacelike instead of null, in the weak accretion limit observers with fixed resolution would still describe their experience as an encounter with an effective shock.  Thus our scenario provides a good first approximation to the experiences of observers who enter weakly-accreting astrophysical black holes at finite times.  While it would be interesting to compute these corrections quantitatively (e.g. by comparing with \cite{accreting}), the fact that in many astrophysical situations the mass accretion rate is far, far less than $M$ per light crossing time\footnote{The light-crossing time of a stellar mass black hole is of order $10^{-4}$ seconds.  Even if such a black hole accreted a solar mass per year, we would have $\dot M \lesssim  10^{-11}$.} suggests that this approximation is quite good indeed.

\section*{Acknowledgements}

DM thanks Patrick Brady, Steve Giddings, Gary Horowitz, Harvey Reall, and Eric Poisson for numerous discussions of black hole interiors over many years.  He in particular thanks Steve Giddings for emphasizing the importance of the late time limit.
AO would like to thank the UCSB physics department for kind hospitality when part of this work was carried out.
D.M. was supported in
part by the National Science Foundation under Grant No PHY08-55415,
and by funds from the University of California.
A.O. was supported in part by the Israel Science Foundation (grant no. 1346/07).

\appendix

\section{Bound on $\Delta E$ in slightly-perturbed RN}

In this Appendix we derive a bound on $\Delta E$, based on Eq. (\ref{eq:dE2f}) and on the inverse-power form of the MP (at the early portion of the CH).

For a {\it generic} situation of a perturbed charged (or spinning) BH, the MP are dominated by the linear term, which is usually a superposition of a term
$\propto v^{-n}$ and a term $\propto u^{-n}$, typically with $n=2l+3$ or possibly  $n=2l+2$, where $l$ is the multipolar number of the MP mode under consideration.
However in our case the situation is different because we are only considering here {\it spherically-symmetric} perturbations, and there are no matter-free spherical MP. Therefore it is only the scalar field which is excited at the linear level. The MP are in turn sourced by the scalar-field energy-momentum tensor (\ref{Tscalar}),
which is quadratic in (derivatives of) $\phi$;
hence the MP will first appear as (and will be dominated by) {\it second-order} perturbations.

Thus, whereas in our case $\phi$ is dominated by terms $\propto v^{-n}$ and $\propto u^{-n}$ (with $n=2l+3=3$), the MP will be dominated by inverse-power terms of overall power $-2n$ rather than $n$.
Restated in other words, the MP will be dominated by a superposition of terms of the form
$c_{jk} v^{-j} u^{-k}$, with non-negative integers $j,k$ satisfying $j+k= 2n=6$,
where $c_{jk}$ are certain constants.
\footnote{To be more specific,
assuming that $\phi$ decays as $v^{-n}$ along the EH, then
perturbation analysis reveals \cite{ORIup} that in the neighborhood of the early portion of the CH $\delta r$ is dominated by
a term $\propto v^{-2n-1}$ plus another term $\propto u^{-2n-1}$,
and $\delta H$ by a term $\propto v^{-n} u^{-n}$
(plus, possibly, additional terms $\propto v^{-2n}$ and $\propto u^{-2n}$, though these two would be more sensitive to the choice of gauge).
Note that as a consequence many of the coefficients $c'_{j'k'}$ (to be defined shortly) actually vanish, though some of them (for example $c'_{n,n+1}$) do not.}

Correspondingly, the term in squared brackets in the RHS of Eq. (\ref{eq:dE2f}) will be dominated by superposition of terms
$c'_{j'k'} v^{-j'} u^{-k'}$ with non-negative integers $j',k'$ satisfying $j'+k'= 2n+1$
(with some new constants $c'_{j'k'}$).

Let us now restrict our attention to the contribution to $d(E^{2})/du$ coming from a single such term $c'_{j'k'} v^{-j'} u^{-k'}$:
\begin{equation}
d(E^{2})_{[j'k']}/du \equiv  c'_{j'k'} v^{-j'} u^{-k'} e^{\kappa (u-v)} .
\label{dEjk}
\end{equation}
We can bound this contribution by
\begin{equation}
 |d(E^{2})_{[j'k']}/du |  < \left [|c'_{j'k'}| (v_{min})^{-j'}e^{-\kappa \, v_{min}} \right] u^{-k'} e^{\kappa u} ,
\label{dEjkBOUND}
\end{equation}
where $v_{min}$ is the minimal value of $v$ in the orbit's section under consideration.
($v_{min}$ is always $\geq v_{eh}$; Later we shall take $v_{min}$ to be the value of $v$ when the orbit approaches the neighborhood of the CH.)

The term in the RHS  of Eq. (\ref{dEjkBOUND}) is an explicit function of $u$.
Consider now its integral
between a certain pair of $u$ values
$u_i,u_f$ satisfying $u_i>u_f\gg M$.
We can now set $v_{min}$ to be $v_i$, namely, the value of $v$ when the orbit approaches $u=u_i$.
Also, since $u\gg M$ throughout the relevant domain,
over the basic exponential time scale $\delta u \sim 1/\kappa$ the factor $u^{-k'}$ does not vary in an appreciable manner, therefore it may be pulled out of the integral. The latter is thus well approximated by just the integrand---namely the RHS of Eq. (\ref{dEjkBOUND})---
divided by $\kappa$ (the error being smaller typically by a factor $\propto M/u$). Thus, denoting the $(j',k')$ contribution to the variation in $E^2$ (accumulated between $u_i$ and $u_f$) by
$\Delta (E^{2})_{[j'k']}$, we find
\[
  \left| \Delta (E^{2})_{[j'k']}  \right|  <
\left [ \kappa^{-1} |c'_{j'k'}| (v_{i})^{-j'}e^{-\kappa v_{i}} \right]
\left [ u^{-k'} e^{\kappa u} \right]_{u_f} ^{u_i} .
\]
Clearly this may be further bounded by the contribution at the upper limit ($u=u_i$), yielding the bound
\begin{equation}
 \left| \Delta (E^{2})_{[j'k']}  \right|  <
 \kappa^{-1} |c'_{j'k'}| (v_{i})^{-j'} (u_i)^{-k'}e^{\kappa (u_i-v_i)}  .
\label{DeltaBOUND}
\end{equation}

The situation which concerns us is the
motion of late-infall observers between two $u=const$ hypersurfaces, $u_1$ to $u_2$ (as described in Sections
\ref{perturbations},\ref{non-linear}).
Our goal here is to show that the late-infall orbits will not cross the CH before approaching $u= u_2$ (for any $u_2\gg M$). We therefore need to bound the variation in $E^{2}$ from the stage where the orbit approaches the neighborhood of the CH, up to $u=u_2$.
Correspondingly, in the above bounds on $ \Delta (E^{2})_{j'k'}$ we should in principle set $u_f=u_2$ (though $u_f$ actually drops out from our final bound), and choose $u_i$ to be the value of $u$ at which the orbit approaches the CH neighborhood.
This notion of ``approaching the CH neighborhood"
is most naturally formulated by means of a certain value of $r_*$, which we denote $r_{*0}$.
That is, a particular orbit is said to having arrived at the CH neighborhood at the point where $r_*=r_{*0}$, and we set our parameters $u_i$ and $v_i$ to be the $(u,v)$ values of that point (this in particular implies $v_i-u_i=2r_{*0}$).
From Eq. (\ref{Delta_r}), $r_{*0}$ is (roughly speaking) the value of $r_*$ at which
$e^{-2 \kappa r_*}$ becomes $\ll 1$.
\footnote{
Certainly there is an arbitrariness in the choice of $r_{*0}$, but it does not affect our final results [e.g. Eqs. (\ref{DeltaBOUNDf}) and (\ref{DeltaEBOUND}) below], which are independent of $r_{*0}$.
Note also that in the analysis here we do not attempt to include modifications in $E$ caused by MP outside the BH, or even inside the BH but not in the CH neighborhood. These ``non-CH" contributions do not concern us here, they can be shown by simpler arguments to decrease as an inverse power of infall time $v_{eh}$.}

Substituting $u_i\to v_i-2r_{*0}$ in Eq. (\ref{DeltaBOUND}),
the exponent at the RHS reads $e^{-2 \kappa r_{*0}}$.
Also, since we are concerned here with the late-infall limit, in evaluating the factor
$(u_i)^{-k'}=(v_i-2r_{*0})^{-k'}$ we can safely ignore $r_{*0}$ compared to $v_i$ (which is $>v_{eh}$),
after which the RHS of Eq. (\ref{DeltaBOUND}) becomes
\begin{equation}
\label{A4}
 \kappa^{-1} |c'_{j'k'}| (v_{i})^{-(j'+k')} e^{-2 \kappa r_{*0}}  .
\end{equation}
Note, however, that the difference between $v_i$ and $v_{eh}$ is, approximately, the quantity $\Delta v$ (defined earlier in Sec. \ref{monotonic}), which is $O(M)$ (and independent of infall time), and hence is $\ll v_{eh}$ in the late-infall limit.
Therefore we can further replace $(v_{i})^{-(j'+k')}$ by $(v_{eh})^{-(j'+k')}$.
Also, since $e^{-2 \kappa r_{*0}} < 1$
(it is in fact $\ll 1$), we are allowed to entirely drop it from our bound.
Recalling that $j'+k'=2n+1$, we obtain the bound on $\Delta (E^{2})_{[j'k']}$ in a much simpler form:
\begin{equation}
 \left| \Delta (E^{2})_{[j'k']}  \right|  <
 \kappa^{-1} \,  |c'_{j'k'}|  \,  (v_{eh})^{-(2n+1)}   .
\label{DeltaBOUND1}
\end{equation}

It remains to sum over the relevant pairs of integers $j'k'$ (recall, these are certain non-negative integers satisfying $j'+k'=2n+1$).
This summation is trivial and we get the bound on $\Delta (E^{2})$ in its final form:
\begin{equation}
 \left| \Delta (E^{2})  \right|  <
C' \, (v_{eh})^{-(2n+1)} =C' \,  (v_{eh})^{-7}   ,
\label{DeltaBOUNDf}
\end{equation}
where $C' = \kappa^{-1} \Sigma_{j'} |c'_{j'k'}|$ (setting $k'=2n+1-j'$).
Obviously the constant $C'$ is independent of $v_{eh}$.

We have thus established that at the late-infall limit, $|\Delta (E^{2})|$ inevitably becomes $\ll E^2$. Equation (\ref{DeltaBOUNDf}) thus also implies that $\Delta E$ (namely the modification in $E$ accumulated in the near-CH region up to some $u_2\gg M$) is bounded for late-infall orbits by\footnote{\label{extend}
For simplicity, the above derivation assumed typical values of the black
hole parameters such that
$\kappa \sim 1/M$;
i.e., the black hole is not too close to extremity.  We note in passing
that a similar argument yields a bound of the same form for any $\kappa > 0$.
}
\begin{equation}
 \left| \Delta E  \right|  <
C \, (v_{eh})^{-(2n+1)} =C \, (v_{eh})^{-7}   ,
\label{DeltaEBOUND}
\end{equation}
where $C=C'/(2E)$.

\end{document}